\documentclass[%
 iop,
 amsmath,amssymb,
preprint,%
 tightenlines, floatfix, nofootinbib
]{revtex4-1}

\usepackage{graphicx}
\usepackage{bm}

\usepackage{array}%
\newcolumntype{C}[1]{>{\centering\let\newline\\\arraybackslash\hspace{0pt}}m{#1}}
\newcolumntype{R}[1]{>{\raggedleft\let\newline\\\arraybackslash\hspace{0pt}}m{#1}}
\newcolumntype{L}[1]{>{\raggedright\let\newline\\\arraybackslash\hspace{0pt}}m{#1}}

\usepackage{longtable}
\usepackage{color} 

\allowdisplaybreaks  

\usepackage[utf8]{inputenc}
\usepackage[T2A,T1]{fontenc} 
\DeclareSymbolFont{cyrillic}{T2A}{cmr}{m}{n}
\DeclareMathSymbol{\de}{\mathalpha}{cyrillic}{228}

\usepackage{esint}  
\usepackage{stmaryrd} 
\usepackage[nointegrals]{wasysym} 

\usepackage{tikz-cd}  
\usetikzlibrary{bending}
\newcommand*\circled[1]{\tikz[baseline=(char.base)]{
    \node[shape=circle,draw,inner sep=0.8pt] (char) {#1};}}

\renewcommand{\thetable}{\arabic{table}}

\renewcommand{\vec}[1]{\boldsymbol {#1}}

\newcommand{\tens}[1]{\boldsymbol{#1}}

\newcommand{\n}{\vec{n}}

\newcommand{\s}{\vec{s}}
\renewcommand{\r}{\vec{r}}
\newcommand{\x}{\vec{x}}
\renewcommand{\u}{\vec{u}}

\newcommand{\A}{\vec{A}}
\newcommand{\B}{\vec{B}}
\renewcommand{\C}{\vec{C}}    
\newcommand{\I}{\vec{I}}   
\newcommand{\V}{\vec{V}}

\newcommand{\X}{\vec{X}}

\newcommand{\0}{\vec{0}}

\newcommand{\D}{\mathcal{D}}
\newcommand{\R}{\mathbb{R}}

\newcommand{\SIm}{\text{m}}
\newcommand{\SIs}{\text{s}}
\newcommand{\SIkg}{\text{kg}}
\newcommand{\SIqty}{\text{qty}}

\renewcommand{\sup}[1]{\textsuperscript{#1}}
\newcommand{\sub}[1]{\textsubscript{#1}}

\newcommand{\G}{\tens{G}}

\newcommand{\wG}{\widetilde{\G}}
\newcommand{\vGamma}{\tens{\Gamma}}
\newcommand{\wGamma}{\widetilde{\tens{\Gamma}}}

\newcommand{\pathcirc}[1]{\raisebox{.5pt}{\textcircled{\raisebox{-.9pt} {#1}}}}

\newcommand{\lvel}{\langle \hspace{-2pt} \langle \hspace{-2pt} \langle}  
\newcommand{\rvel}{\rangle \hspace{-2pt} \rangle \hspace{-2pt} \rangle} 
\newcommand{\lvol}{[ \hspace{-1pt} [ \hspace{-1pt}  [} 
\newcommand{\rvol}{] \hspace{-1pt} ] \hspace{-1pt}  ]}  

\newcommand{\dV}{dV}  
\newcommand{\dU}{dU}  

\newcommand{\volint}{\iiint\limits_{\Omega(\u,t)} \dV}
\newcommand{\velint}{\iiint\limits_{\D(\x,t)} \dU}
\newcommand{\volintalt}{\iiint\limits_{\Omega(t)} \dV}
\newcommand{\velintalt}{\iiint\limits_{\D(t)} \dU}

\newcommand{\Usp}{\mathcal{U}}
\newcommand{\Vsp}{\mathcal{V}}
\newcommand{\Qsp}{\mathcal{Q}}
\newcommand{\msp}{\mathsf{m}}
\newcommand{\Msp}{\mathsf{M}}

\newcommand{\lexp}{[}
\newcommand{\rexp}{]}

\newcommand{\vpartial}{\vec{\partial}}

\DeclareMathOperator{\vol}{vol}

\newcommand{\delf}{\vec{\blacktriangledown}}  
\newcommand{\vecunit}{\vec{d}} %
\newcommand{\comp}{\llcorner} 

\newcommand{\ins}[1]{\breve{#1}}
\newcommand{\loc}[1]{\underline{#1}}

\begin{document}


\title[New Classes of Conservation Laws]{New Classes of Conservation Laws Based on Generalized Fluid Densities and Reynolds Transport Theorems}

\author{Robert K. Niven}
\email{Email r.niven@adfa.edu.au}
\affiliation{School of Engineering and Information Technology, The University of New South Wales, Northcott Drive, Canberra ACT 2600, Australia.}

\date{\today}

\begin{abstract}
The Reynolds transport theorem occupies a central place in fluid dynamics, providing a generalized integral conservation equation for the transport of any conserved quantity within a fluid, and connected to its corresponding differential equation.
Recently, a new generalized framework was presented for this theorem, enabling parametric transformations between positions on a manifold or in a generalized coordinate space, exploiting the underlying multivariate Lie symmetries associated with a conserved quantity. 
We examine the implications of this framework for fluid flow systems, within an Eulerian position-velocity (phase space) description. 
The analysis invokes a hierarchy of five probability density functions, which by convolution are used to define five fluid densities and generalized densities appropriate for different spaces. 
We obtain 11 formulations of the generalized Reynolds transport theorem for different choices of the coordinate space, parameter space and density, only the first of which is known. 
These are used to generate 11 tables of integral and differential conservation laws applicable to these systems, for eight important conserved quantities (fluid mass, species mass, linear momentum, angular momentum, energy, charge, entropy and probability). 
These substantially expand the set of conservation laws for the analysis of fluid flow and dynamical systems.
\end{abstract}

\maketitle

\section{\label{Intro}Introduction} 

Near the end of his distinguished career, Osborne Reynolds presented what is now called the {\it Reynolds transport theorem}: a generalized conservation equation for the transport of a conserved quantity within a body of fluid (the {\it domain} or {\it fluid volume}) as it moves through a prescribed region of space (the {\it control volume}) \cite{Reynolds_1903}.  This provides a universal formulation for the construction of integral conservation equations for any conserved quantity, and can be connected to corresponding differential equations for these quantities \citep[e.g.,][]{White_1986, Munson_etal_2010, deGroot_M_1984, Bird_etal_2006, White_2006, Durst_2008}. Such conservation laws -- founded on the paradigm of the field or Eulerian description of fluid flow -- provide the basis for most theoretical and numerical analyses of flow systems.
Extensions of the Reynolds transport theorem have been presented for 
moving and smoothly-deforming control volumes \cite{White_1986, Munson_etal_2010, Dvorkin_Goldschmidt_2005}, 
domains with fixed or moving discontinuities \cite{Truesdell_Toupin_1960, Dvorkin_Goldschmidt_2005, Myers_2015}, 
irregular and rough domains \cite{Seguin_Fried_2014, Seguin_etal_2014, Falach_Segev_2014}, 
two-dimensional domains \cite{Gurtin_etal_1989, Ochoa-Tapia_etal_1993, Slattery_Sagis_Oh_2007, Fosdick_Tang_2009, Lidstrom_2011}, 
and differentiable manifolds using the formalism of exterior calculus \cite{Flanders_1973, Lee_2009, Frankel_2013, Harrison_2015}. The Reynolds transport theorem is also a special case of the Helmholtz transport theorem, for flow through an open and moving surface \cite{Tai_1992}. 

Traditionally, the Reynolds transport theorem is viewed exclusively as a continuous one-parameter (temporal) mapping of the density of a conserved quantity in geometric space, along the pathlines described by a time-dependent velocity vector field. Indeed, the above formulations all conform to this tradition.  However, Flanders \cite{Flanders_1973} interpreted the theorem more broadly as a generalization of the Leibniz rule for differentiation of an integral, rather than simply a construct of fluid mechanics. It is therefore far more general and powerful than the traditional interpretation might suggest.  
Using this insight, a generalized framework for the Reynolds transport theorem has recently been presented \cite{Niven_etal_2020, Niven_etal_2020b}, based on continuous multiparametric mappings of a differential form on a manifold -- or of a density within a generalized coordinate space -- connected by the maximal integral curves described by a vector or tensor field. 
This extends the traditional interpretation to encompass new {\it transformation theorems}, which exploit previously unreported multiparametric continuous (Lie) symmetries associated with a conserved quantity in the space considered.
These can be used, for example, to connect different positions in a velocity space connected by a velocity gradient tensor field, different positions in a spectral space connected by a velocity-wavenumber tensor field, or different positions in a velocity and chemical species space connected by velocity and concentration gradient tensor fields \cite{Niven_etal_2020, Niven_etal_2020b}. The generalized framework also yields many new forms of the Liouville equation for the conservation of probability in different spaces, and of the Perron-Frobenius and Koopman operators for the extrapolation of probability or observable densities in such systems \cite{Niven_etal_2020}.

The aim of this work is to present a consolidated treatment of the generalized Reynolds transport theorem -- and consequential integral and differential conservation laws -- arising from an Eulerian position-velocity (phase space) description of fluid flow systems.  The analysis commences in \S\ref{sect:description} with a detailed discussion of this description, and of the properties of the position and velocity domains for several well-known classes of fluid flow systems. This leads in \S\ref{sect:densities} to a hierarchy of densities, starting in \S\ref{sect:pdfs} with five probability density functions (pdfs), which are defined and in which their commutative relations are examined. These are used in \S\ref{sect:fluid_densities} to define five fluid densities, of which only the volumetric density $\rho$ has previously been in common use for the analysis of fluid flow systems, while the remaining four have many similarities to other densities (such as the phase space density) used in other branches of physics.  The fluid densities are formally defined from the pdfs by mathematical convolutions, for which the definitions and philosophical implications are discussed in Appendix \ref{sect:ApxA} and Appendix \ref{sect:ApxB}. The fluid densities are then used in \S\ref{sect:gen_densities} to construct corresponding generalized densities for any conserved quantity. In \S\ref{sect:ReTT}, we present the generalized framework for the Reynolds transport theorem, in both exterior calculus and vector calculus formulations.  The latter is then applied in \S\ref{sect:ex_sys} to give 11 formulations of the Reynolds transport theorem arising from the position-velocity description, for different choices of the coordinate space, parameter space and density. Of these, only the first was given by Reynolds \cite{Reynolds_1903}. 
These are used to generate 11 tables of integral and differential conservation equations for these systems, for the eight conserved quantities commonly considered in fluid mechanics (fluid mass, species mass, linear momentum, angular momentum, energy, charge, entropy and probability). 
For some systems, it is possible to extract partial differential equations from the Reynolds transport theorem -- as explained in Appendix \ref{sect:ApxC} -- while in others, it is necessary to extract Lie differential equations, containing the Lie derivative of a volume form in the domain. 
The analyses provide a considerable assortment of new conservation laws for the analysis of fluid flow systems.

In the following sections, the mathematical notation is defined when first used and is also listed in Appendix \ref{sect:ApxD}.


\section{\label{sect:description}The Position-Velocity Description and Domains} 

In continuum mechanics, fluid flow systems are commonly examined using the {\it Eulerian description}, in which each local property of the fluid is specified as a function of Cartesian coordinates $\x=[x,y,z]^\top \in \Omega \subset \R^3$ and time $t \in \R$ as the fluid moves past, where $\Omega$ is a three-dimensional geometric space and $^\top$ is the transpose.  Thus for example, fluid mechanists commonly consider the three-dimensional velocity $\u(\x,t)$, the volumetric mass density $\rho(\x,t)$ and the volumetric mass concentration $\rho_c(\x,t)$ of the $c$th chemical species within the Eulerian description. 
We here consider an alternative {\it position-velocity continuum description} of a dynamical system based on Eulerian position-velocity coordinates, in which each local property of a fluid is specified as a function of the instantaneous fluid velocity $\u = [u,v,w]^\top \in \D \subset \R^3$, position $\x \in \Omega \subset \R^3$ and time $t \in \R$ as the fluid moves past, where $\D$ is a three-dimensional velocity space.  This treatment -- analogous to the phase space description used in many branches of physics -- has the advantage of explicitly incorporating the velocity dependence of physical quantities, significantly extending the breadth of physical quantities that can be considered, and the scope and fidelity of the analyses that can be conducted.

\begin{figure}[t]
\begin{center}
 \begin{picture}(140,150)
 \put(0,80){ \includegraphics[height=70pt]{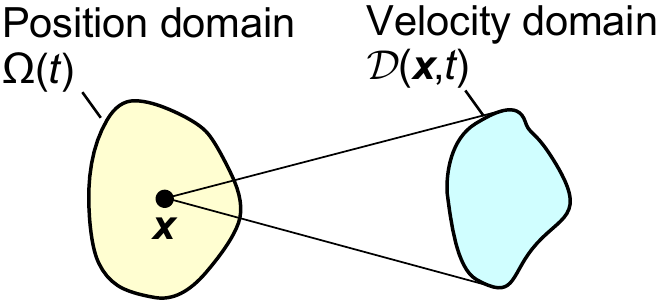} }
 \put(0,80){\small (a)} 
 \put(0,0){ \includegraphics[height=70pt]{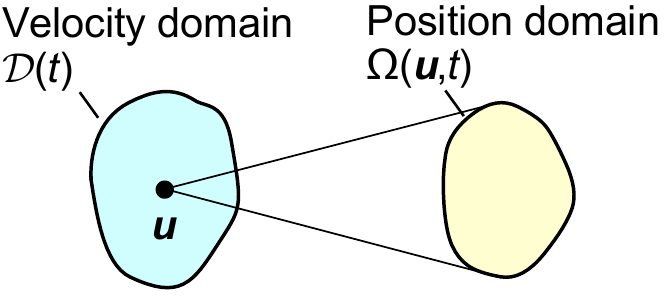}}
 \put(0,0){\small (b)}
 \end{picture}
\end{center}
\caption{Schematic diagrams of the mapping between domains $\Omega$ and $\D$ in (a) the geometric representation, and (b) the velocimetric representation.}
\label{fig:domains}
\end{figure}

We now examine two alternative representations of the domains $\Omega$ and $\D$:
\newcounter{LcountA}
\begin{list}{(\alph{LcountA})}{\usecounter{LcountA} \topsep 3pt \itemsep 3pt \parsep 0pt \leftmargin 18pt \rightmargin 0pt \listparindent 0pt \itemindent 0pt}
\item The {\it geometric representation} -- the usual physical viewpoint -- in which $\D(\x,t)$ is a function of position and time, and $\Omega(t)$ is a function of time. In this perspective, as illustrated in Fig.\ \ref{fig:domains}(a), there exists a map between each position $\x \in \Omega(t)$ and an entire velocity space $\D(\x,t)$, consisting of all possible velocities for this position and time.
\item The {\it velocimetric representation} -- an alternative viewpoint -- in which $\Omega(\u,t)$ is a function of velocity and time, and $\D(t)$ is a function of time. In this perspective, as illustrated in Fig.\ \ref{fig:domains}(b), there exists a map between each velocity $\u \in \D(t)$ and an entire position space $\Omega(\u,t)$, consisting of all possible positions for this velocity and time.
\end{list}
In principle, the set of all ordered triples $(\u,\x,t)$ for a given flow system can be mapped into either of these representations, hence $\D(\x,t) \subseteq \D(t)$ for all $\x \in \Omega(t)$ and $\Omega(\u,t) \subseteq \Omega(t)$ for all $\u \in \D(t)$.

Examining the velocity domain $\D(t)$ or $\D(\x,t)$, we make two important assertions. Firstly, we expect $\D(t)$ or $\D(\x,t)$ to be continuous, since for most flows, it is physically impossible for a local velocity to change from $\u_1$ to $\u_2$ without passing through all intermediate velocities $\u_1 < \u < \u_2$, however fleetingly. The main exception to this rule are flows which cross a regime threshold, e.g., from laminar to turbulent flow, or subsonic to supersonic flow, leading to two or more disjoint but internally continuous velocity domains. Secondly, the velocity cannot be infinite (positive or negative) for any physically realizable flow, as this would require local velocities of infinite kinetic energy. In consequence, for all flows $\D(t)$ or $\D(\x,t)$ should be bounded, and for many flows they will also be closed. We note that integration of the velocity over $\R^3$ serves as a useful assumption for many calculations, but this is simply an approximation that cannot be manifested physically.

In consequence, each domain in Fig.\ \ref{fig:domains} is drawn as compact and simply connected, a useful starting assumption for  turbulent flow systems, but there will be many exceptions.  For most flows involving a continuous fluid volume with no  velocity discontinuities, the primary domains $\Omega(t)$ and $\D(t)$ and the subsidiary domain $\D(\x,t)$ should be compact and simply connected, but some subsidiary domains $\Omega(\u,t)$ may consist of multiple disjoint regions, each associated with a different location (or set of locations) within the flow. By breaking the original problem into smaller coupled flows, or by a judicious choice of coordinate system, it may be possible to isolate or unite these regions. 
In flows with different flow regimes, the velocity domains $\D(t)$ and $\D(\x,t)$ will consist of disjoint regions. In such cases, it may be possible to isolate each region using a dimensionless discriminator (such as a Reynolds, Mach or Froude number). 
For some flows, for example the turbulent boundary layer, there are long-standing arguments over whether the overall position domain $\Omega(t)$ can be considered compact, due to a lack of boundedness or closedness, but despite this the subsidiary position domains $\Omega(\u,t)$ and the overall and subsidiary velocity domains $\D(t)$ and $\D(\x,t)$ will very likely be bounded and may also be closed.
For homogeneous isotropic turbulence -- admittedly a highly idealized flow -- each subsidiary domain $\D(\x,t)$ or $\Omega(\u,t)$ must be independent respectively of position or velocity, and so the two representations will collapse to give two separable domains $\Omega(t)$ and $\D(t)$. 
In laminar flows, each subsidiary velocity domain $\D(\x,t)$ can be idealized as a single point (in reality, allowing for fluctuations, a small region), while each subsidiary position domain $\Omega(\u,t)$ will consist of single or multiple disconnected points (or small regions).

\section{\label{sect:densities}A Hierarchy of Densities} 

\subsection{\label{sect:pdfs}Probability Density Functions and Domains} 

We can now define the primary probability density functions (pdfs) that underlie fluid flow systems, and indeed can be used to define the physical densities used in such systems. 
Writing the nonnegative real line as $\R^{+}_0$, the two descriptions give rise to the following five pdfs\footnote{We here use standard pdf notation with a common symbol $p$, leading to a mixed signature and functional notation.}:
\newcounter{Lcount1}
\begin{list}{(\alph{Lcount1})}{\usecounter{Lcount1} \topsep 3pt \itemsep 3pt \parsep 0pt \leftmargin 18pt \rightmargin 0pt \listparindent 0pt \itemindent 0pt}
\item A volumetric pdf $p(\x|t): \Omega \times \R \to \R^{+}_0$ [SI units: m\sup{-3}];
\item A velocimetric pdf $p(\u|t): \D \times \R \to \R^{+}_0$  [(m s\sup{-1})\sup{-3}]; 
\item A velocimetric and volumetric (phase space) pdf $p(\u,\x|t): \D \times \Omega \times \R \to \R^{+}_0$ [m\sup{-3} (m s\sup{-1})\sup{-3}];
\item A conditional velocimetric (ensemble) pdf $p(\u|\x, t): \D \times \Omega \times \R \to \R^{+}_0$ [(m s\sup{-1})\sup{-3}]; and 
\item A conditional volumetric pdf $p(\x |\u, t): \Omega \times \D \times \R \to \R^{+}_0$ [m\sup{-3}];
\end{list}
where the solidus $|$ is the conditional probability symbol, with the conditions listed to the right.

\begin{figure}[!t]
\setlength{\unitlength}{1pt}
\centering
     	\begin{tikzpicture}[font=\normalsize]
	        \node[rectangle] (O) at (3.5, 1.8) {$\mathop{1} \limits_{[-]}$};
		\node[rectangle] (A) at (1.75,0) {$\mathop{p(\x | t)} \limits_{[\SIm^{-3}]}$};
		\node[rectangle] (B) at (5.25,0) {$ \mathop{p(\u | t)} \limits_{[(\SIm \, \SIs^{-1})^{-3}]}$};
		\node[rectangle] (C) at (3.5,-1.8) {$\mathop{p(\u,\x|t)} \limits_{[\SIm^{-3} \, (\SIm \, \SIs^{-1})^{-3}]}$ };
		\node[rectangle] (P) at (0, -1.8) {$\mathop{p(\u |\x, t)} \limits_{[(\SIm \, \SIs^{-1})^{-3}]}$};
		\node[rectangle] (Q) at (7,-1.8) {$\mathop{p(\x | \u, t)} \limits_{[\SIm^{-3}]}$}; 
		\draw [->] (B) edge node[rotate=0,above] {\hspace{30pt} denom.} (Q);
		\draw [->] (C) edge [bend right=25] node[rotate=0,below] {numer.} (Q);
		\draw [->] (A) edge node[rotate=0,above] {denom. \hspace{30pt} } (P);
		\draw [->] (C) edge [bend left=25] node[rotate=0,below] {numer.} (P);
                 \draw [->] (P) edge [bend left=70] node[rotate=0,above] {$\velint \hspace{10pt}$} (O);
                 \draw [->] (Q) edge [bend right=70] node[rotate=0,above] {$\hspace{25pt} \volint$} (O);
		%
		\draw[->] (C) -- node[rotate=0,below, near end] {$\velint \hspace{30pt}$} (A);
		\draw[->] (C) -- node[rotate=0,below, near end] {$\hspace{30pt} \volint$} (B);
		\draw[->] (A) -- node[rotate=0,above, near start] {$\volintalt \hspace{15pt}$} (O);
		\draw[->] (B) -- node[rotate=0,above, near start] {$\hspace{30pt} \velintalt$} (O);
                %
	\end{tikzpicture}
\caption{Relational diagram between the pdfs defined in this study.}
\label{fig:rel_diag_pdfs}
\end{figure}
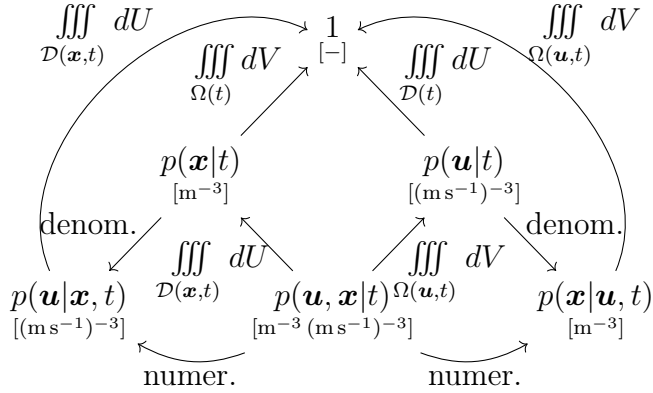

Using the notation $\dV=dxdydz=d^3 \x$ for an infinitesimal volume element and $\dU=dudvdw = d^3 \u$ for an infinitesimal three-dimensional velocity element, the five pdfs will satisfy the following nine relations:
\begin{align}
1 &= \iiint\nolimits_{\Omega(t)} p(\x|t) \, \dV 
\label{eq:pdf_rel_pxsubt}
\\
1 &= \iiint\nolimits_{\D(t)} p(\u|t) \, \dU
\\
\begin{split}
1 &= \iiint\nolimits_{\Omega(t)}  \iiint\nolimits_{\D(\x,t)} p(\u,\x|t) \, \dU \, \dV
\\&= \iiint\nolimits_{\D(t)}  \iiint\nolimits_{\Omega(\u, t)}  p(\u,\x|t) \, \dV \, \dU
\end{split}
\label{eq:pdf_rel_puxsubt}
\\
p(\x |t) &= \iiint\nolimits_{\D(\x,t)}p (\u,\x |t) \, \dU
\label{eq:pdf_puxsubt_int_left}
\\
p(\u | t) &= \iiint\nolimits_{\Omega(\u,t)} p(\u,\x | t) \, \dV
\label{eq:pdf_puxsubt_int_right}
\\
p(\u | \x, t) &= \frac{p(\u,\x|t)}{p(\x | t)} = \frac{p(\u,\x|t)}{\iiint\nolimits_{\D(\x,t)}p (\u,\x |t) \dU}
\label{eq:pdf_pusubxt_def}
\\
p(\x | \u, t) &= \frac{p(\u,\x|t)}{p(\u | t)} = \frac{p(\u,\x|t)}{\iiint\nolimits_{\Omega(\u,t)} p(\u,\x | t) \dV}
\label{eq:pdf_pxsubut_def}
\\
1 &= \iiint\nolimits_{\D(\x,t)} p(\u | \x, t) \, \dU
\label{eq:pdf_norm_pusubxt}
  \\
1 &= \iiint\nolimits_{\Omega(\u,t)} p(\x | \u, t) \, \dV 
\label{eq:pdf_norm_pxsubut}
\end{align}
in which Eq.\ \eqref{eq:pdf_rel_puxsubt} uses an inside-out order of integration. 
The connections between the five pdfs, and the roles of the different domain representations, are illustrated in the relational diagram in Fig.\ \ref{fig:rel_diag_pdfs}. As evident, the geometric representation is used in the integration paths on the left-hand side of Fig.\ \ref{fig:rel_diag_pdfs}, while the velocimetric representation is required on the right-hand side. 

We recognize $p(\x|t)$ as the common probabilistic descriptor for fluid flow systems, forming the basis of the fluid mechanics formulations of the Liouville and Fokker-Planck equations \cite{Liouville_1838, Risken_1984, Lutzen_1990, Ehrendorfer_2003, Pottier_2010}, and allied to the volumetric density $\rho(\x,t)$. Physically (albeit imprecisely \cite{Feller_1966}), $p(\x|t) \, \dV$ can be interpreted as the probability that a fluid element is situated in the position between $\x$ and $\x + d\x$, at the time between $t$ and $t + dt$.  Furthermore, $p(\u |\x, t) \, \dU$ can be interpreted as the probability that a fluid element has a velocity between $\u$ and $\u + d\u$, at the position between $\x$ and $\x + d\x$ and time between $t$ and $t + dt$. We thus recognize $p(\u | \x,t)$ -- typically but incorrectly written as $p(\u)$ -- as the basis of the ensemble mean often used in the Reynolds-averaged Navier-Stokes (RANS) equations, and of the single-position correlation functions of turbulent fluid mechanics \cite{Batchelor_1967, Monin_Yaglom_1971a, Hinze_1975, Pope_2000, Davidson_2004}.  

Both the above pdfs and also $p(\u |t)$ and $p(\x | \u,t)$ can be derived from the Eulerian phase space pdf $p(\u,\x |t)$. Physically, $p(\u,\x |t) \, \dU \, \dV$ can be interpreted as the joint probability of a fluid element having a velocity between $\u$ and $\u + d\u$ and position between $\x$ and $\x + d\x$, at the time between $t$ and $t + dt$. As evident from Eqs.\ \eqref{eq:pdf_puxsubt_int_left}-\eqref{eq:pdf_puxsubt_int_right} and Fig.\ \ref{fig:rel_diag_pdfs}, marginalization of $p(\u,\x |t)$ over the velocity space or geometric space respectively yields $p(\x |t)$ or $p(\u | t)$, while its quotient Eq.\ \eqref{eq:pdf_pusubxt_def} or Eq.\ \eqref{eq:pdf_pxsubut_def} with respect to each of these pdfs yields the conditional pdfs $p(\u | \x,t)$ and $p(\x|\u,t)$. 

In addition to the pdfs in Fig.\ \ref{fig:rel_diag_pdfs}, it is possible to consider joint pdfs with respect to time, including $p(\x,t)$, $p(\u,t)$ and $p(\u,\x,t)$. These require normalization over a time interval in addition to their volume and/or velocity domain(s). Such pdfs are closely associated with path-based formulations for the description of entire histories of events \cite{Jaynes_1980, Gonzalez_etal_2013, Dixit_etal_2017, Ghosh_etal_2020}, and are not considered further here.

\subsection{\label{sect:fluid_densities}Fluid Densities}

The above five pdfs can be used to define corresponding fluid densities, four of which are not commonly used for the analysis of fluid flow systems\footnote{We here revert to standard signature and functional notation.}:
\newcounter{Lcount2}
\begin{list}{(\alph{Lcount2})}{\usecounter{Lcount2} \topsep 3pt \itemsep 3pt \parsep 0pt \leftmargin 18pt \rightmargin 0pt \listparindent 0pt \itemindent 0pt}
\item A volumetric fluid density $\rho: \Omega \times \R \to \R^{+}_0$, $(\x,t) \mapsto \rho(\x,t)$ [kg m\sup{-3}]; 
\item A velocimetric fluid density $\de: \D \times \R \to \R^{+}_0$, $(\u,t) \mapsto \de(\u,t)$  [kg (m s\sup{-1})\sup{-3}]; 
\item A velocimetric and volumetric (phase space) fluid density $\zeta: \D \times \Omega \times \R \to \R^{+}_0$, $(\u,\x,t) \mapsto \zeta(\u,\x,t)$ [kg m\sup{-3} (m s\sup{-1})\sup{-3}]; 
\item A conditional velocimetric (ensemble) fluid density $\eta: \D \times \Omega \times \R \to \R^{+}_0$, $(\u,\x,t) \mapsto \eta(\u,\x,t)$ [kg (m s\sup{-1})\sup{-3}]; and 
\item A conditional volumetric fluid density $\xi: \D \times \Omega \times \R \to \R^{+}_0$, $(\u,\x,t) \mapsto \xi(\u,\x,t)$ [kg m\sup{-3}]; 
\end{list}
The symbols for the last four densities, including the Cyrillic ``de'' character (from the Russian transliteration of ``density''), have been chosen to not conflict with the most common notation of fluid mechanics. 

\begin{figure}[t]
\setlength{\unitlength}{1pt}
\centering
	 \begin{tikzpicture}[font=\normalsize]
	        \node[rectangle] (O) at (3.5, 1.8) {$\mathop{M} \limits_{[\SIkg]}$};
		\node[rectangle] (A) at (1.75,0) {$\mathop{\rho(\x, t)} \limits_{[\SIkg \, \SIm^{-3}]}$};
		\node[rectangle] (B) at (5.25,0) {$ \mathop{\de(\u, t)} \limits_{[\SIkg \, (\SIm \, \SIs^{-1})^{-3}]}$};
		\node[rectangle] (C) at (3.5,-1.8) {$\mathop{\zeta(\u,\x, t)} \limits_{[\SIkg \, \SIm^{-3} \, (\SIm \, \SIs^{-1})^{-3}]}$ };
		\node[rectangle] (P) at (0, -1.8)  {$\mathop{\eta(\u ,\x, t)} \limits_{[\SIkg \,(\SIm \, \SIs^{-1})^{-3}]}$};
		\node[rectangle] (Q) at (7,-1.8) {$\mathop{\xi(\u, \x, t)} \limits_{[\SIkg \, \SIm^{-3}]}$};  
		\draw [->] (B) edge node[rotate=0,above] {\hspace{30pt} denom.} (Q);
		\draw [->] (C) edge [bend right=25] node[rotate=0,below] {numer.} (Q);
		\draw [->] (A) edge node[rotate=0,above] {denom. \hspace{30pt} } (P);
		\draw [->] (C) edge [bend left=25] node[rotate=0,below] {numer.}  (P);
                 \draw [<->] (P) edge [bend left=70] node[rotate=0,above] {$\velint \hspace{10pt}$} node[rotate=0,below, near start] {numer.} (O);
                 \draw [<->] (Q) edge [bend right=70] node[rotate=0,above] {$\hspace{25pt} \volint$} node[rotate=0,below, near start] {numer.} (O);
		%
		\draw[->] (C) -- node[rotate=0,below, near end] {$\velint \hspace{30pt}$} (A);
		\draw[->] (C) -- node[rotate=0,below, near end] {$\hspace{30pt} \volint$} (B);
		\draw[->] (A) -- node[rotate=0,above, near start] {$\volintalt \hspace{15pt}$} (O);
		\draw[->] (B) -- node[rotate=0,above, near start] {$\hspace{30pt} \velintalt$} (O);
                %
	\end{tikzpicture}
\caption{Relational diagram between the fluid densities defined in this study.}
\label{fig:rel_diag_densities}
\end{figure}
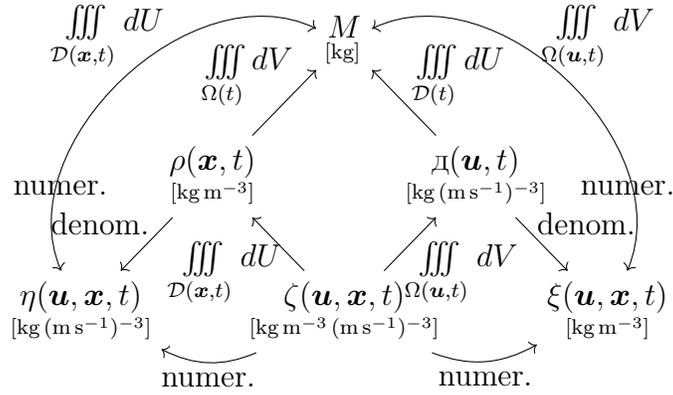

For analysis, consider a fluid volume containing the fluid mass $M$, which in the absence of sources or sinks of fluid will be constant in time\footnote{For control volumes containing sources or sinks of fluid mass, a more comprehensive analysis is needed.}. Firstly, we see that the five fluid densities must satisfy the following nine relations:
\begin{alignat}{2}
M 
&= \iiint\nolimits_{\Omega(t)} \rho(\x,t) \, \dV
\label{eq:dens_rho_norm}
\\
M 
&= \iiint\nolimits_{\D(t)} \de(\u,t) \,  \dU
\label{eq:dens_de_norm}
\\
\begin{split}
M 
&= \iiint\nolimits_{\Omega(t)}  \iiint\nolimits_{\D(\x,t)} \zeta(\u,\x,t) \, \dU \, \dV
\\&= \iiint\nolimits_{\D(t)}  \iiint\nolimits_{\Omega(\u, t)}  \zeta(\u,\x,t) \, \dV \, \dU
\end{split}
\label{eq:dens_zeta_int_double}
\\
\rho(\x,t) 
&= \iiint\nolimits_{\D(\x,t)} \zeta(\u,\x,t) \, \dU
\label{eq:dens_zeta_int_left}
\\
\de(\u, t) 
&= \iiint\nolimits_{\Omega(\u,t)} \zeta(\u,\x,t) \, \dV
\label{eq:dens_zeta_int_right}
  \\
\eta(\u, \x, t) &= \frac{\zeta(\u,\x,t) \, M}{\rho(\x, t)} = \frac{\zeta(\u,\x,t) \, M}{\iiint\nolimits_{\D(\x,t)} \zeta(\u,\x,t) \dU}
\label{eq:dens_eta_def}
\\
\xi(\u, \x, t) &= \frac{\zeta(\u,\x,t) \, M}{\de(\u, t)} = \frac{\zeta(\u,\x,t) \, M}{\iiint\nolimits_{\Omega(\u,t)} \zeta(\u,\x,t) \dV}
\label{eq:dens_xi_def}
\\
M &= \iiint\nolimits_{\D(\x,t)} \eta(\u, \x, t) \, \dU
\label{eq:dens_eta_norm}
  \\
M &= \iiint\nolimits_{\Omega(\u, t)} \xi(\u, \x, t) \, \dV 
\label{eq:dens_xi_norm}
\end{alignat}
Clearly these are analogs of Eqs.\ \eqref{eq:pdf_rel_pxsubt}-\eqref{eq:pdf_norm_pxsubut}, modified only by the introduction of $M$ into the definitions of $\eta$ and $\xi$. Note that the time dependence is lost in the integrations in Eqs.\ \eqref{eq:dens_rho_norm}-\eqref{eq:dens_zeta_int_double} and Eqs.\ \eqref{eq:dens_eta_norm}-\eqref{eq:dens_xi_norm}; furthermore Eqs.\ \eqref{eq:dens_eta_norm}-\eqref{eq:dens_xi_norm} also lose their dependence respectively on $\x$ or $\u$. 
The connections between the different fluid densities are illustrated in the relational diagram in Fig.\ \ref{fig:rel_diag_densities}.

Secondly, following a rich line of research \citep[e.g.,][]{Matheron_1965, Marle_1967, Anderson_Jackson_1967, Whitaker_1967a, Whitaker_1969, Bachmat_1972, Gray_Lee_1977, Hassanizadeh_Gray_1979a, Narasimhan_1980, Ene_1981, Cushman_1982, Cushman_1983b, Baveye_Sposito_1984, Cushman_1984, Cushman_1985, Bear_Bachmat_1991, Gray_etal_1993, Quintard_Whitaker_1993, Chen_1994, Grau_Cantero_1994, Cushman_1997, Whitaker_1999, Gray_Miller_2005b,  Wood_2013, Gray_Miller_2013, Pokrajac_deLemos_2015, Davit_Quintard_2017, Takatsu_2017}, the fluid density $\rho$ at each time $t$ is commonly defined by integration over a small fluid volume $\Vsp$ -- or equivalently a small fluid mass $\msp$ -- for which there are two interpretations. In the common viewpoint, $\Vsp$ must be sufficiently large to enable the fluid to be considered a continuum  \citep[e.g.,][]{Whitaker_1967a, Whitaker_1969, Bachmat_1972, Gray_Lee_1977, Bear_Bachmat_1991, Gray_etal_1993}. Thus in single phase systems, it must provide a ``microscopic'' scale large enough to average out the molecular phenomena, while for multiphase systems it may need to be larger than the dominant ``macroscopic'' scales \cite{Whitaker_1969, Quintard_Whitaker_1993, Hassanizadeh_Gray_1979a}. However, in the contrary ``relativist'' viewpoint, $\Vsp$ is not considered a property of the continuum, but is simply a characteristic of the measurement scale \cite{Baveye_Sposito_1984, Cushman_1984, Chen_1994, Cushman_1997}.
The analyses here encompass both viewpoints; we also adopt a small velocimetric domain $\Usp$ for velocity averaging. We adopt the local Cartesian position coordinates $\r = [r_x, r_y, r_z]^\top \in \Omega \subset \R^3$ and velocity coordinates $\s= [s_u, s_v, s_w]^\top \in \D \subset \R^3$ aligned with their corresponding global coordinates, and with their origin respectively at $\x$ or $\u$. The small domains $\Vsp$ and $\Usp$ are functions respectively of $\x$ and $\u$, as well as time, and must conform to the two domain representations introduced earlier, bringing additional dependencies on $\r$ and/or $\s$ into the definition of $\zeta$. The five fluid densities can then be defined from their underlying pdfs by the following convolutions: 
\begin{align}
\begin{split}
\lexp \rho \rexp &(\x,t) 
= \int\nolimits_{\msp(\x,t)} p(\r|t) \, dm(\x+\r, t)  \\
&= \iiint\nolimits_{\Vsp(\x,t)} p(\r|t) \, \rho(\x + \r, t) \, \dV(\r, t)
\label{eq:pxsubt_rho}
\end{split}
\\
\begin{split}
\langle \de \rangle &(\u,t)
 = \int\nolimits_{\msp(\u,t)} p(\s|t) \, dm(\u+\s, t) \\
 &= \iiint\nolimits_{\Usp(\u,t)} p(\s|t) \, \de(\u + \s,t) \, \dU(\s, t)
\end{split}
\\
\begin{split}
\lexp \langle \zeta \rangle \rexp &(\u,\x,t) 
 = \int\nolimits_{\msp(\u,\x,t)} p(\s,\r|t) \, dm(\u+\s, \x+\r, t) \\
&= \iiint\nolimits_{\Vsp(\x,t)}  \iiint\nolimits_{\Usp(\u,\x+\r,t)} p(\s,\r|t) \, \zeta(\u+\s,\x+\r,t) \, 
\\ & \hspace{80pt} \times \dU(\s, \x+ \r, t) \, \dV(\s, \r, t)
\\&= \iiint\nolimits_{\Usp(\u,t)}  \iiint\nolimits_{\Vsp(\u+\s,\x,t)} p(\s,\r|t) \, \zeta(\u+\s,\x+\r,t) \, 
\\ & \hspace{80pt} \times \dV(\u+\s, \r, t) \, \dU(\s, \r, t)
\end{split}
\\
\begin{split}
\langle \eta \rangle &(\u, \x, t)  
 = \int\nolimits_{\msp(\u,\x,t)} p(\s|\x, t) \, dm(\u+\s, \x, t) \\
&= \iiint\nolimits_{\Usp(\u,\x,t)} p(\s | \x, t) \, \eta(\u + \s, \x, t) \, \dU(\s, \x, t)
\end{split}
  \\
\begin{split}
\lexp \xi \rexp &(\u, \x, t)
 = \int\nolimits_{\msp(\u,\x,t)} p(\r|\u, t) \, dm(\u, \x+\r, t) \\
  &= \iiint\nolimits_{\Vsp(\u,\x,t)} p(\r | \u, t) \, \xi(\u, \x+\r, t) \, \dV(\u, \r, t)
\label{eq:puxsubt_xi}
\end{split}
\end{align}
where $\lexp \cdot \rexp$ is a local volumetric expectation, $\langle \cdot \rangle$ is a local velocimetric expectation, and $dm$ is an infinitesimal element of fluid mass. 
A more detailed discussion of these definitions is given in Appendix \ref{sect:ApxA}, while the philosophical implications of the use of pdfs to define physical variables are explored in Appendix \ref{sect:ApxB}. 
Note that if each pdf is assumed uniformly distributed over its domain, each expected fluid density reduces to the product of its underlying pdf and the fluid mass, as would be obtained from dimensional considerations. 

In this study we adopt the mass integrals in Eqs.\ \eqref{eq:pxsubt_rho}-\eqref{eq:puxsubt_xi} as the primary definitions of the fluid densities, since the volumetric and velocimetric integrals require knowledge of point density terms that need to be defined. For convenience the expectation notations used in Eqs.\ \eqref{eq:pxsubt_rho}-\eqref{eq:puxsubt_xi} are now dropped.

In accordance with their underlying pdfs, we see that $\eta$, $\zeta$, $\xi$ and $\rho$ are local densities, i.e., they apply to each point in the flow field, while $\eta$, $\zeta$, $\xi$ and $\de$ are velocimetric densities, i.e., they apply to fluid elements with a specified velocity. The well-known volumetric fluid density $\rho$ represents the fluid mass per unit volume carried by a fluid element in the position between $\x$ and $\x + d\x$, at the time between $t$ and $t + dt$. By definition, the phase space density $\zeta$ represents the fluid mass carried per unit volume and velocity space by a fluid element of velocity between $\u$ and $\u + d\u$ and position between $\x$ and $\x + d\x$, at the time between $t$ and $dt$. The conditional ensemble density $\eta$ then represents the fluid mass per unit velocity space carried by a fluid element with a velocity between $\u$ and $\u + d\u$, at the position between $\x$ and $\x + d\x$ and time between $t$ and $t + dt$, while the conditional density $\xi$ represents the fluid mass per unit volume carried by a fluid element in the position between $\x$ and $\x + d\x$, with a velocity between $\u$ and $\u + d\u$ and time between $t$ and $t + dt$.  The distinctions between $\xi$, $\eta$ and $\zeta$ are therefore quite subtle,  but since they arise from separate underlying pdfs, each density is mathematically well-defined.

In contrast, the density $\de$ and its underlying pdf $p(\u |t)$ are non-local densities, applying over the entire fluid volume $\Omega(t)$. Physically, $\de$ represents the fluid mass per unit of velocity space carried through the entire fluid volume by fluid elements of velocity between $\u$ and $\u + d\u$, at the time between $t$ and $dt$. This is a very strange, aggregated density, but nonetheless both it and its underlying pdf are well-defined. 

We see that Fig.\ \ref{fig:rel_diag_densities} is an analog of Fig.\ \ref{fig:rel_diag_pdfs}, exhibiting the same vertical symmetry, with the geometric representation used on the left-hand side (integration paths in Eqs.\ \eqref{eq:dens_rho_norm}, \eqref{eq:dens_zeta_int_left} and \eqref{eq:dens_eta_norm}), and the velocimetric representation used on the right-hand side (integration paths in Eqs.\ \eqref{eq:dens_de_norm}, \eqref{eq:dens_zeta_int_right} and \eqref{eq:dens_xi_norm}). For example, the integration of $\zeta(\u,\x,t)$ to $\de(\u,t)$ requires a geometric domain consisting of all possible positions $\x$ corresponding to the velocity $\u$ and time $t$, hence $\Omega(\u,t)$. The integration of $\xi(\u,\x,t)$ also requires this velocimetric representation of the geometric domain. 

\subsection{\label{sect:gen_densities}Generalized Densities} 

Based on the above definitions, we can now define five generalized densities of a conserved quantity (an extensive variable) carried by a fluid flow:
\newcounter{Lcount3}
\begin{list}{(\alph{Lcount3})}{\usecounter{Lcount3} \topsep 3pt \itemsep 3pt \parsep 0pt \leftmargin 18pt \rightmargin 0pt \listparindent 0pt \itemindent 0pt}
\item Volumetric densities $\alpha(\x,t)$ [qty m\sup{-3}];
\item Velocimetric densities $\beta(\u,t)$ [qty (m s\sup{-1})\sup{-3}];
\item Phase space densities $\varphi(\u,\x,t)$ [qty m\sup{-3} (m s\sup{-1})\sup{-3}];
\item Conditional velocimetric (ensemble) densities $\theta(\u,\x,t)$ [qty (m s\sup{-1})\sup{-3}]; and
\item Conditional volumetric densities $\epsilon(\u,\x,t)$ [qty m\sup{-3}];
\end{list}
where ``$\text{qty}$'' denotes the units of the conserved quantity. These can be defined by the following relations:
\begin{align}
\alpha(\x,t)  &= \rho(\x,t) \, \loc{\alpha}(\x,t) 
\label{eq:a_def}
\\
\beta(\u,t) &= \de(\u,t)  \, \ins{\beta}(\u,t) 
\label{eq:b_def}
\\
\varphi(\u,\x,t) &= \zeta(\u,\x,t)  \, \loc{\ins{\varphi}}(\u,\x,t) 
\label{eq:f_def}
\\
\theta(\u,\x,t) &= \eta(\u,\x,t)  \, \loc{\ins{\theta}}(\u,\x,t) 
\label{eq:g_def}
\\
\epsilon(\u,\x,t) &= \xi(\u,\x,t)  \, \loc{\ins{\epsilon}}(\u,\x,t) 
\label{eq:h_def}
\end{align}
where $\loc{\alpha}$, $\ins{\beta}$, $\loc{\ins{\varphi}}$, $\loc{\ins{\theta}}$ and $\loc{\ins{\epsilon}}$ [qty kg\sup{-1}] are specific densities, representing the quantity carried per unit fluid mass. For precision, these are designated by their functional dependencies; e.g., for the specific energy, $\loc{e}$ indicates a local density (a function of $\x$), $\ins{e}$ indicates a velocity-distinct density (a function of $\u$), and $\loc{\ins{e}}$ indicates dependence on both $\u$ and $\x$ (hence $\loc{e}=\iiint\nolimits_{\D(\x,t)} \loc{\ins{e}} \, \dU$ and $\ins{e} = \iiint\nolimits_{\Omega(t)}  \loc{\ins{e}} \, \dV$). However, the specific momentum density $\u$ need not carry these designations -- being already velocity-dependent -- provided that care is taken over its dependence on position. 

\begin{figure}[t]
\setlength{\unitlength}{1pt}
\centering
	 \begin{tikzpicture}[font=\normalsize]
	        \node[rectangle] (O) at (3.5, 1.8) {$\mathop{Q(t)} \limits_{[\SIqty]}$};
		\node[rectangle] (A) at (1.75,0) {$\mathop{\alpha(\x, t)} \limits_{[\SIqty \, \SIm^{-3}]}$};
		\node[rectangle] (B) at (5.25,0) {$ \mathop{\beta(\u, t)} \limits_{[\SIqty \, (\SIm \, \SIs^{-1})^{-3}]}$};
		\node[rectangle] (C) at (3.5,-1.8) {$\mathop{\varphi(\u,\x, t)} \limits_{[\SIqty \, \SIm^{-3} \, (\SIm \, \SIs^{-1})^{-3}]}$ };
		\node[rectangle] (P) at (0, -1.8)  {$\mathop{\theta(\u ,\x, t)} \limits_{[\SIqty \,(\SIm \, \SIs^{-1})^{-3}]}$};
		\node[rectangle] (Q) at (7,-1.8) {$\mathop{\epsilon(\u, \x, t)} \limits_{[\SIqty \, \SIm^{-3}]}$};  
		\draw [->] (B) edge node[rotate=0,above] {\hspace{30pt} denom.} (Q);
		\draw [->] (C) edge [bend right=25] node[rotate=0,below] {numer.} (Q);
		\draw [->] (A) edge node[rotate=0,above] {denom. \hspace{30pt} } (P);
		\draw [->] (C) edge [bend left=25] node[rotate=0,below] {numer.}  (P);
                 \draw [<->] (P) edge [bend left=70] node[rotate=0,above] {$\velint \hspace{10pt}$} node[rotate=0,below, near start] {numer.} node[rotate=0,below, red, font=\small] {\hspace{3pt} \circled{5}} (O);
                 \draw [<->] (Q) edge [bend right=70] node[rotate=0,above] {$\hspace{25pt} \volint$} node[rotate=0,below, near start] {numer.} node[rotate=0,below, red, font=\small] {\circled{6} \hspace{10pt}} (O);
		%
		\draw[->] (C) -- node[rotate=0,below, near end] {$\velint \hspace{30pt}$} node[rotate=0,above, red, font=\small] {\hspace{3pt} \circled{3}} (A);
		\draw[->] (C) -- node[rotate=0,below, near end] {$\hspace{30pt} \volint$} node[rotate=0,above, red, font=\small] {\circled{4} \hspace{10pt}} (B);
		\draw[->] (A) -- node[rotate=0,above, near start] {$\volintalt \hspace{15pt}$} node[rotate=0,below, red, font=\small] {\hspace{3pt} \circled{1}} (O);
		\draw[->] (B) -- node[rotate=0,above, near start] {$\hspace{30pt} \velintalt$} node[rotate=0,below, red, font=\small] {\circled{2} \hspace{10pt}}  (O);
                %
	\end{tikzpicture}
\caption{{Relational diagram between the generalized densities defined in this study} (the integration paths are numbered in red).}
\label{fig:rel_diag_gen_densities}
\end{figure}
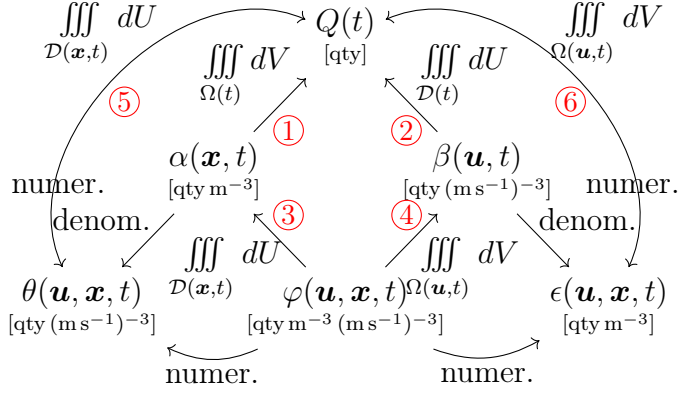

We define $Q(t)$ to be the total conserved quantity (of any type) in the volumetric domain, which due to the action of sources or sinks of the conserved quantity will be a function of time $t$. The generalized densities will satisfy the following nine integral relations:
\begin{alignat}{2}
Q(t)
&= \iiint\nolimits_{\Omega(t)} \alpha(\x,t) \, \dV
\label{eq:dens_alpha_norm}
\\
Q(t) 
&= \iiint\nolimits_{\D(t)} \beta(\u,t) \,  \dU
\label{eq:dens_beta_norm}
\\
\begin{split}
Q(t) 
&= \iiint\nolimits_{\Omega(t)}  \iiint\nolimits_{\D(\x,t)} \varphi(\u,\x,t) \, \dU \, \dV
\\&= \iiint\nolimits_{\D(t)}  \iiint\nolimits_{\Omega(\u, t)}  \varphi(\u,\x,t) \, \dV \, \dU
\end{split}
\label{eq:dens_varphi_int_double}
\\
\alpha(\x,t) 
&= \iiint\nolimits_{\D(\x,t)} \varphi(\u,\x,t) \, \dU
\label{eq:dens_varphi_int_left}
\\
\beta(\u, t) 
&= \iiint\nolimits_{\Omega(\u,t)} \varphi(\u,\x,t) \, \dV
\label{eq:dens_varphi_int_right}
  \\
\theta(\u, \x, t) &= \frac{\varphi(\u,\x,t) \, Q(t)}{\alpha(\x, t)} = \frac{\varphi(\u,\x,t) \, Q(t)}{\iiint\nolimits_{\D(\x,t)} \varphi(\u,\x,t) \dU}
\label{eq:dens_theta_def}
\\
\epsilon(\u, \x, t) &= \frac{\varphi(\u,\x,t) \, Q(t)}{\beta(\u, t)} = \frac{\varphi(\u,\x,t) \, Q(t)}{\iiint\nolimits_{\Omega(\u,t)} \varphi(\u,\x,t) \dV}
\label{eq:dens_epsilon_def}
\\
Q(t) &= \iiint\nolimits_{\D(\x,t)} \theta(\u, \x, t) \, \dU
\label{eq:dens_theta_norm}
  \\
Q(t) &= \iiint\nolimits_{\Omega(\u, t)} \epsilon(\u, \x, t) \, \dV 
\label{eq:dens_epsilon_norm}
\end{alignat}
Clearly these are analogs of Eqs.\ \eqref{eq:dens_rho_norm}-\eqref{eq:dens_xi_norm}, containing $Q(t)$ rather than $M$. The connections between generalized densities are shown in the relational diagram in Fig.\ \ref{fig:rel_diag_gen_densities} (compare Figs.\ \ref{fig:rel_diag_pdfs} and \ref{fig:rel_diag_densities}). 
By synthesis of Eqs.\ \eqref{eq:pxsubt_rho}-\eqref{eq:puxsubt_xi} and \eqref{eq:a_def}-\eqref{eq:h_def}, we can also extract the relationship between each generalized density and its underlying pdf (see discussion in Appendix \ref{sect:ApxA}).

Finally, we note that all formulations in this section are functions of time, and in general of both position and velocity. These can be simplified to give quantities and domains that are not functions of time, position or velocity, leading to a considerable assortment of reduced mathematical formulations.

\section{\label{sect:ReTT}Generalized Formulations of Conservation Equations}

\subsection{\label{sect:ext_calc_Re_tr}Exterior Calculus Formulations} 

We now examine generalized forms of the Reynolds transport theorem, which can be interpreted more broadly as {\it transformation theorems}, i.e., they provide a continuous mapping within a domain, described by the maximal integral curves of a vector or tensor field defined with respect to a parameter space. Traditionally, this is used to define a mapping between positions connected by a velocity field, parameterized by time, to give the usual Reynolds transport theorem \cite{Reynolds_1903}. However, as has been shown \cite{Niven_etal_2020, Niven_etal_2020b}, this is not the only example. 

For maximum generality we adopt an exterior calculus formulation \citep[e.g.,][]{Kobayashi_Nomizu_1963, Guggenheimer_1963, Cartan_1970, Flanders_1973, Lovelock_Rund_1989, Olver_1993, Lee_2009, Torres_del_Castillo_2012, Bachman_2012, Frankel_2013, Sjamaar_2017}. Consider an $r$-dimensional oriented compact submanifold $\Omega^r$ within an $n$-dimensional orientable differentiable manifold $\Msp^n$, described using a patchwork of local coordinate systems. Let $\V$ be a vector or tensor field in $\Msp^n$, a function of the local coordinates and parameterized by the $m$-dimensional parameter vector $\C$, but not a function of $\C$. The components $C_c$ of $\C$ are assumed orthogonal. The field trajectories (tangent bundles) of $\V$ define the continuous multivariate map (``flow'') $\phi^{\C}: \Msp^n \to \Msp^n$ such that $\V = (\partial \phi^{\C}/\partial \C)^\top$. This is linear and invertible, and can be used to map the entire submanifold. 
If $\omega^r$ is an $r$-form representing a conserved quantity, its integral over $\Omega^r$ can be proved to satisfy \cite{Niven_etal_2020}:
\begin{align}
\begin{split}
d \int\limits_{\Omega(\C)}  \omega^r 
&=  \biggl[ \int\limits_{\Omega(\C)} \mathcal{L}_{\V}^{(\C)}  \omega^r \biggr] \cdot {d\C} 
\\&=  \biggl[ \int\limits_{\Omega(\C)}  \, i_{\V}^{(\C)}  \, d  \omega^r + {\oint\limits_{\partial \Omega(\C)}  }  \,   i_{\V}^{(\C)}  \, \omega^r 
\biggr]  \cdot {d\C} 
\\
&=   \biggl[  \int\limits_{\Omega(\C)}  
 i_{\V}^{(\C)}  \, d \omega^r + d  ( i_{\V}^{(\C)} \, \omega^r  )
\biggr] \cdot  {d\C} 
\end{split}
\label{eq:Re_tr_diffform}
\end{align}
where $d$ is the exterior derivative, $\partial \Omega$ is the submanifold boundary, ``$\cdot$'' is the dot product, $\mathcal{L}_{\V}^{(\C)}$ is a multivariate Lie derivative with respect to $\V$ over parameters $\C$, and $i_{\V}^{(\C)}$ is a multivariate interior product with respect to $\V$ over parameters $\C$. The multivariate operators provide vector extensions of their usual one-parameter definitions in exterior calculus \cite{Niven_etal_2020}, while the last step in Eq.\ \eqref{eq:Re_tr_diffform} invokes Stokes' theorem, imposing a regularity condition on $\omega^r$ in the submanifold $\Omega$. 
For $\C=t$, Eq.\ \eqref{eq:Re_tr_diffform} reduces to the one-parameter exterior calculus formulation of the Reynolds transport theorem \cite{Frankel_2013}. 

If the vector or tensor field $\V$ is also a function of $\C$, the problem can be analyzed by augmenting the manifold with the parameter space, to define the flow $\hat{\phi}^C: \Msp^n \times \R^m \to \Msp^n \times \R^m$ based on the augmented field $\V \comp \C = (\partial \hat{\phi}^{\C}/\partial \C)^\top$, where $\comp$ denotes this composition \cite{Niven_etal_2020}. This is the $(n+m) \times m$ tensor given by $\V \comp \C = [\begin{smallmatrix} \V \\ \I_m \end{smallmatrix}]$, where $\I_m$ is the identity matrix of size $m$. Applying Eq.\ \eqref{eq:Re_tr_diffform} based on $\V \comp \C$, this simplifies to the extended theorem \cite{Niven_etal_2020}: 
\begin{align}
\begin{split}
\hat{d} \int\limits_{\Omega(\C)}  \omega^r 
&=  \biggl[ \int\limits_{\Omega(\C)} \mathcal{L}_{\V \comp \C}^{(\C)}  \, \omega^r \biggr] \cdot {d\C} 
\\&=   \biggl[  \int\limits_{\Omega(\C)}  
\vpartial_{\C} \omega^r +  i_{\V}^{(\C)}  \, {d} \omega^r + {d} ( i_{\V}^{(\C)} \, \omega^r  )
\biggr] \cdot  {d\C} 
\end{split}
\raisetag{55pt}
\label{eq:Re_tr_diffform_ext}
\end{align}
where $\hat{d}$ is the extended exterior derivative based on the augmented coordinates, and $\vpartial_{\C}$ is a vector partial derivative operator with respect to the components of $\C$. For Cartesian parameters $\C$, $\vpartial_{\C}=\nabla_{\C}$, while for $\X=\x$, $\C=t$ and $\V=\u$, Eq.\ \eqref{eq:Re_tr_diffform_ext} reduces to an exterior calculus formulation based on a time-varying velocity field $\u(t)$ \cite{Flanders_1973, Frankel_2013}.

We can now extract the Lie differential equations applicable to each differential form in $\Omega$, respectively from Eqs.\ \eqref{eq:Re_tr_diffform} and \eqref{eq:Re_tr_diffform_ext}:
\begin{align}
\begin{split}
\mathcal{L}_{\V}^{(\C)}  \omega^r 
&=    
 i_{\V}^{(\C)}  \, d \omega^r + d  ( i_{\V}^{(\C)} \, \omega^r  )
\end{split}
\label{eq:DE_diffform}
\\
\begin{split}
\mathcal{L}_{\V \comp \C}^{(\C)}  \omega^r 
&=    
\vpartial_{\C} \omega^r +  i_{\V}^{(\C)}  \, {d} \omega^r + {d} ( i_{\V}^{(\C)} \, \omega^r  )
\end{split}
\label{eq:DE_diffform_ex}
\end{align}
The former is the multivariate extension of Cartan's relation of exterior calculus \citep[c.f.,][]{Lovelock_Rund_1989, Lee_2009, Frankel_2013}, while the latter provides an extended form for $\V=\V(\C)$.

\subsection{\label{sect:vect_calc_Re_tr}Vector Calculus Formulations} 

Eqs.\ \eqref{eq:Re_tr_diffform}-\eqref{eq:Re_tr_diffform_ext} provide quite general equations applicable to submanifolds of any dimension in a manifold. 
For a system with global coordinates, these can be simplified to give a generalized parametric Reynolds transport theorem. Consider an $n$-dimensional domain $\Omega$ within an $n$-dimensional space $\Msp$, described by the global Cartesian coordinates $\X$. Let $\V = (\partial \X/\partial \C)^\top = (\nabla_{\C} \X)^\top$ be a vector or tensor field in $\Msp$, using the $\partial (\to)/\partial (\downarrow)$ vector derivative convention, in general with $\V$ a function of $\C$. Let $\omega^n$ be an $n$-dimensional volume form defined by:
\begin{align}
\omega 
= \psi \, dX_1 \wedge ... \wedge dX_n = \psi \, \vol^n_{\X}
\label{eq:nform_psi}
\end{align}
where $\psi (\X,\C)$ is the density of a conserved quantity, $\wedge$ is the wedge product and $\vol^n_{\X}$ is the volume of an infinitesimal $n$-dimensional parallelopiped spanned by the cotangents to $\X$. We assume $\psi$ is continuous and continuously differentiable with respect to $\X$ and $\C$ throughout $\Omega$. It can be shown that Eq.\ \eqref{eq:Re_tr_diffform_ext} then reduces to \cite{Niven_etal_2020}:
\begin{align}
\begin{split}
d &\int\limits_{\Omega(\C)} \psi \, d^n \X 
= \biggl[ \nabla_{\C} \int\limits_{\Omega(\C)} \psi \, d^n \X \biggr] \cdot d\C
\\&=  \biggl[  \int\limits_{\Omega(\C)}  \nabla_{\C} \psi  \, d^n \X +{\oint\limits_{\partial \Omega(\C)} }  \; \psi \, \V \cdot d^{n-1} \X  \biggr] \cdot {d\C} 
\\&
=    \biggl[ \int\limits_{\Omega(\C)}  \bigl[ \nabla_{\C} \psi +  \nabla_{\X} \cdot \bigl(  \psi \, \V  \bigr) \bigr] d^n \X  \biggr] \cdot {d\C},
\end{split}
\label{eq:Re_tr_gen}
\raisetag{95pt}
\end{align}
where $d$ is now the differential operator, 
$\nabla_{\X}$ is the $n$-dimensional nabla operator with respect to $\X$, 
$\nabla_{\C}$ is the $m$-dimensional nabla operator with respect to $\C$, 
$d^n \X$ is a volume element in $\Omega$, 
and 
$d^{n-1} \X$ is a directed area element with outward unit normal on the boundary $\partial \Omega$. 
For consistency with the derivative convention, the divergence in Eq.\ \eqref{eq:Re_tr_gen} is defined by 
$ \nabla_{\X} \cdot (  \psi \, \V  ) =  [\nabla_{\X}^\top (  \psi \, \V  )]^\top$. 

We emphasize that Eq.\ \eqref{eq:Re_tr_gen} applies to a smoothly varying domain $\Omega$, with coordinates $\X$ measured with respect to a fixed frame of reference. Further extensions to consider moving and smoothly deforming frames of reference (control domains) \citep[c.f.,][]{White_1986, Munson_etal_2010, Niven_etal_2020}, domains with jump discontinuities \cite{Truesdell_Toupin_1960, Dvorkin_Goldschmidt_2005}, or irregular and fragmenting domains \cite{Seguin_Fried_2014, Seguin_etal_2014, Falach_Segev_2014} can also be derived. 

We now combine the exterior calculus (Eq.\ \eqref{eq:Re_tr_diffform_ext}) and vector calculus formulations (Eq.\ \eqref{eq:Re_tr_gen}) formulations, to provide a generalized Lie differential equation applicable to each volume form in $\Omega$.  Using the notation in the second part of Eq.\ \eqref{eq:nform_psi}, we rewrite the first part of Eq.\ \eqref{eq:Re_tr_diffform_ext}:  
\begin{align}
\hat{d} \int\limits_{\Omega(\C)} \psi \, \vol^n_{\X}
=  \biggl[ \int\limits_{\Omega(\C)} \mathcal{L}_{\V \comp \C}^{(\C)}  (\psi \, \vol^n_{\X})  \biggr] \cdot {d\C} 
\label{eq:Re_tr_gen_Lie}
\end{align}
Since the exterior derivative of the integral is equivalent to its differential, and integration of $\psi$ with respect to $\vol^n_{\X}$ is identical to integration over $d^n \X$, we also have:
\begin{align}
\hat{d} \int\limits_{\Omega(\C)} \psi \, \vol^n_{\X}
= d \int\limits_{\Omega(\C)} \psi \, d^n \X 
\label{eq:Re_tr_gen_Lie2}
\end{align}
Combining Eqs.\ \eqref{eq:Re_tr_gen}, \eqref{eq:Re_tr_gen_Lie} and \eqref{eq:Re_tr_gen_Lie2} and equating integrands gives, for each volume form in $\Omega$: 
\begin{align}
 \mathcal{L}_{\V \comp \C}^{(\C)}  (\psi \, \vol^n_{\X})
 =    \bigl[ \nabla_{\C} \psi +  \nabla_{\X} \cdot \bigl(  \psi \, \V  \bigr) \bigr] d^n \X  
\label{eq:DE_gen_Lie}
\end{align}
This invokes the fundamental lemma of the calculus of variations, thus imposing a regularity assumption on $\psi$ and $\V$ within $\Omega$. The generalized Reynolds transport theorem in Eq.\ \eqref{eq:Re_tr_gen} thus yields a Lie differential equation \eqref{eq:DE_gen_Lie}, defined using the augmented field $\V \comp \C$. In some flow systems, this can be simplified to give a partial differential equation applicable to each local element $d^n \X$. 

\section{\label{sect:ex_sys}Example Flow Systems}
In the following sections we consider eleven case study flow systems, for different choices of the coordinates $\X$,  parameters $\C$, vector or tensor field $\V$ and generalized density $\psi$ in Eqs.\ \eqref{eq:Re_tr_gen} and \eqref{eq:DE_gen_Lie}. 
These follow the six integration paths labelled in red in Fig.\ \ref{fig:rel_diag_gen_densities}, with some choices examined for both time-independent and time-dependent systems. 
To establish a consistent nomenclature, each system is named using its $\X-\C$ coordinates and its selected density.
The components of $\X$ and $\C$ are assumed orthogonal. 
In all cases we consider $\V=\V(\C)$, but report only the {\it intrinsic} equations, with $\V$ measured with respect to a fixed frame of reference. 
We also consider only smoothly-varying fields within a compact and simply-connected domain. 

\subsection{ \label{sect:exA} Volumetric-Temporal Formulation (Density $\alpha$)} 

We first consider the well-known volumetric-temporal formulation of the Reynolds transport theorem \cite{Reynolds_1903}, based on the geometric position space $\Omega$ with Cartesian coordinates $\X=\x$, time parameter $\C=t$, intrinsic velocity vector field $\V=\u(\x,t):=\partial \x / \partial t$ and generalized density $\psi=\alpha(\x,t)$, defined from the fluid density $\rho(\x,t)$. This follows integration path \pathcirc{1} in Fig.\ \ref{fig:rel_diag_gen_densities}.
From Eq.\ \eqref{eq:Re_tr_gen} and the definition of $\alpha$ in Eq.\ \eqref{eq:dens_alpha_norm}, we obtain the well-known equation:
\begin{align}
\begin{split}
\frac{dQ}{dt} 
= \frac{DQ}{Dt} 
&= \frac{d}{dt} \iiint\limits_{\Omega(t)} \alpha \, \dV 
=    \iiint\limits_{\Omega(t)}  \frac{\partial \alpha}{\partial t}  \, \dV + \oiint\limits_{\partial \Omega(t)} \alpha  \u \cdot \n dA 
\\  
&=   \iiint\limits_{\Omega(t)}  \biggl[ \frac{\partial \alpha }{\partial t}   +  \nabla_{\x} \cdot (\alpha  \u ) \biggr ] \dV  
\end{split}
\label{eq:Re_tr_vol-temp-alpha}
\raisetag{30pt}
\end{align}
using $dV=d^3 \x$ and $\n dA = d\A = d^2 \x$, where $\n$ is the outward unit normal, $dA$ is an area element and $d\A$ is a directed area element. Eq.\ \eqref{eq:Re_tr_vol-temp-alpha} is commonly written in terms of the substantial derivative $D/Dt={\partial}/{\partial t} + \u \cdot \nabla_{\x}$, expressing the transport of the conserved quantity with the fluid volume.

To extract the differential equation, we require a local form of the left-hand side of Eq.\ \eqref{eq:Re_tr_vol-temp-alpha}. From the continuity equation:
\begin{align}
\begin{split}
0
&= \frac{\partial \rho  }{\partial t}   +  \nabla_{\x} \cdot (\rho  \u ) 
\end{split}
\label{eq:DE_tr_vol-temp-alpha_cty}
\end{align}
a simple manipulation of Eq.\ \eqref{eq:Re_tr_vol-temp-alpha} using the local specific density $\loc{\alpha}$ in Eq.\ \eqref{eq:a_def} gives:
\begin{align}
\begin{split}
 \rho \frac{D \loc{\alpha}}{Dt}
&= \frac{\partial (\rho \loc{\alpha}) }{\partial t}   +  \nabla_{\x} \cdot (\rho \loc{\alpha} \u ) 
\end{split}
\label{eq:DE_tr_vol-temp-alpha}
\end{align}
Further details of this derivation are given in \cite{Hutter_Johnk_2004} and Appendix \ref{sect:ApxC}.

The integral and differential equations \eqref{eq:Re_tr_vol-temp-alpha} and \eqref{eq:DE_tr_vol-temp-alpha} provide generalized forms of the standard conservation laws of fluid mechanics.  In these equations, the left-hand terms are generally used as placeholders for any source-sink terms or driving forces for the conserved quantity to enter the fluid volume or differential fluid element. These equations, for the seven common conserved quantities (fluid mass, chemical species mass, linear momentum, angular momentum, energy, charge and entropy), are listed in Table \ref{tab:Re_tr_vol-temp-alpha_table}. These contain typical source-sink terms for the left-hand side of each equation, here labelled to indicate the fluid volume FV$=\Omega$; more comprehensive versions for coupled phenomena can also be derived \citep[e.g.,][]{deGroot_M_1984, White_1986, Kuiken_1994, Spurk_1997, Demirel_2002, Hutter_Johnk_2004, Bird_etal_2006, Munson_etal_2010}. 
For reference, the symbols used in this work are listed in Table \ref{tab:nomen} in Appendix \ref{sect:ApxD} (some minor overlaps of symbols could not be avoided). To enable dimensional comparisons, the SI units of each integral and differential equation are also tabulated.

To the seven conservation laws, we can also add an eighth, by assigning $\alpha(\x,t)$ to its underlying pdf $p(\x | t)$\footnote{Strictly, following the probabilistic averaging method of Appendix \ref{sect:ApxA}, this is achieved by assigning $\alpha(\x+\r,t)=p(\x+\r|t)$ in Eqs.\ \eqref{eq:a_def_prob_vol}-\eqref {eq:a_point_def} or $\loc{\alpha}(\x+r, t)=1$ [kg\sup{-1}] in Eq.\ \eqref{eq:a_def_prob_qty}, producing a probabilistic convolution.}. In this case, by normalization $\iiint\nolimits_{\Omega(t)} p(\x | t) \, \dV =1$, the left-hand term of the Reynolds transport theorem in Eq.\ \eqref{eq:Re_tr_vol-temp-alpha} vanishes. In consequence, for a compactly supported continuous and continuously differentiable pdf $p(\x|t)$ \cite{Weinstock_1952, Gelfand_Fomin_1963} we obtain the differential equation \cite{Niven_etal_2020}:
\begin{align}
0 =   \frac{\partial p(\x | t) }{\partial t}   +  \nabla_{\x} \cdot (p(\x | t) \,  \u ) 
\label{eq:Liouville_vol-temp-alpha}
\end{align}
This is the {\it Liouville equation} of fluid mechanics \cite{Liouville_1838, Lutzen_1990, Ehrendorfer_2003, Pottier_2010}. This and its integral form are included in Table \ref{tab:Re_tr_vol-temp-alpha_table}. As will be shown, other Liouville equations based on different pdfs can be derived for other representations.

\subsection{Velocimetric-Temporal Formulation (Density $\beta$)} \label{sect:exB} 

Now consider a velocimetric-temporal formulation of the Reynolds transport theorem, based on the Eulerian velocity space $\D$ with Cartesian velocity coordinates $\X=\u$, time parameter $\C=t$, local acceleration vector field $\V=\dot{\u}(\u,t):=\partial \u / \partial t$ and generalized density $\psi=\beta(\u,t)$, defined from the velocimetric fluid density $\de(\u,t)$. This follows integration path \pathcirc{2} in Fig.\ \ref{fig:rel_diag_gen_densities}.
Recall that this requires $\dot{\u}$, $\beta$ and $\de$ to be defined in the velocimetric representation, for fluid elements of velocity $\u$ aggregated over all positions. 
From Eq.\ \eqref{eq:Re_tr_gen} and the definition of $\beta$ in Eq.\ \eqref{eq:dens_beta_norm} \cite{Niven_etal_2020b}:
\begin{align}
\begin{split}
\frac{dQ}{dt} 
&= \frac{d}{dt} \iiint\limits_{\D(t)} \beta \, \dU
=    \iiint\limits_{\D(t)}  \frac{\partial \beta}{\partial t}  \, \dU + \oiint\limits_{\partial \D(t)} \beta  \dot{\u} \cdot \n_B \, dB
\\  
&=   \iiint\limits_{\D(t)}  \biggl[ \frac{\partial \beta }{\partial t}   +  \nabla_{\u} \cdot (\beta  \dot{\u}) \biggr ] \dU 
\end{split}
\label{eq:Re_tr_vel-temp-beta}
\raisetag{45pt}
\end{align}
using $dU=d^3 \u$ and $\n_B \, dB = d\vec{B} = d^2 \u$, where $\n_B$ is the outward unit normal, $dB$ is a velocimetric boundary element and $d\vec{B}$ is a directed velocimetric boundary element. 
Using the previous manipulation (Appendix \ref{sect:ApxC}), we recover a velocimetric analog of the continuity equation at each velocity:
\begin{align}
\begin{split}
 0
&= \frac{\partial \de }{\partial t}   +  \nabla_{\u} \cdot (\de \dot{\u}) 
\end{split}
\label{eq:DE_tr_vel-temp-beta_cty}
\end{align}
and in general the differential equation based on $\ins{\beta}$:
\begin{align}
\begin{split}
 \de \frac{d \ins{\beta}}{dt}
&= \frac{\partial (\de \ins{\beta}) }{\partial t}   +  \nabla_{\u} \cdot (\de \ins{\beta} \dot{\u} ) 
\end{split}
\label{eq:DE_tr_vel-temp-beta}
\end{align}

The conservation laws derived from Eqs.\ \eqref{eq:Re_tr_vel-temp-beta} and \eqref{eq:DE_tr_vel-temp-beta} are listed with their SI units  in Table \ref{tab:Re_tr_vel-temp-beta_table}. 
Note that from Eqs.\ \eqref{eq:Re_tr_vol-temp-alpha} and \eqref{eq:Re_tr_vel-temp-beta} (see also Fig.\ \ref{fig:rel_diag_gen_densities}), each integral on path \pathcirc{2} is equal to the same rate of change as its volumetric counterpart on path \pathcirc{1} (see Table \ref{tab:Re_tr_vol-temp-alpha_table}), but is now identified with the velocity volume VV$=\D$. 
In contrast, the differential equations are localized by velocity rather than position, and so the source-sink terms are not immediately identifiable, but nonetheless can be written in the form of Eq.\ \eqref{eq:DE_tr_vel-temp-beta}, providing convenient placeholders for all source-sink terms.  The corresponding  integral and differential {Liouville} equations, obtained by assigning $\beta(\u,t)=p(\u | t)$, are also listed in Table \ref{tab:Re_tr_vel-temp-beta_table}.

\subsection{Velocivolumetric-Temporal Formulation (Density $\varphi$)}  \label{sect:exC}

Now consider a velocivolumetric-temporal formulation of the Reynolds transport theorem, based on the Eulerian velocity-position space $\D \times \Omega$ with six-dimensional Cartesian coordinates $\X=[\begin{smallmatrix} \u \\ \x \end{smallmatrix}]$, time parameter $\C=t$, composite vector field $\V=[\begin{smallmatrix} \dot{\u} \\ \u \end{smallmatrix}](\u,\x,t):=\partial [\begin{smallmatrix} \u \\ \x \end{smallmatrix}] / \partial t$ and generalized density $\psi=\varphi(\u,\x,t)$, defined from the phase space density $\zeta(\u,\x,t)$. This follows the double integration path \pathcirc{3}-\pathcirc{1} or \pathcirc{4}-\pathcirc{2} in Fig.\ \ref{fig:rel_diag_gen_densities}. For path \pathcirc{3}-\pathcirc{1}, from Eq.\ \eqref{eq:Re_tr_gen} and the definition of $\varphi$ in Eq.\ \eqref{eq:dens_varphi_int_double}:
\begin{align}
\begin{split}
&\frac{dQ}{dt} 
= \frac{d}{dt} \iiint\limits_{\Omega(t)} \iiint\limits_{\D(\x,t)}  \varphi \, \dU \dV 
\\
&=    \iiint\limits_{\Omega(t)} \iiint\limits_{\D(\x,t)}  \frac{\partial \varphi}{\partial t}  \, \dU \dV 
+ \oiint\limits_{\partial \Omega(t)} \oiint\limits_{\partial \D(\x,t)} \varphi \Biggl[ \begin{matrix} \dot{\u} \\ \u \end{matrix} \Biggr]   \cdot \Biggl[ \begin{matrix} \n_B \\ \n \end{matrix} \Biggr] dB \, dA
\\  
&=   \iiint\limits_{\Omega(t)}  \iiint\limits_{\D(\x,t)}  \biggl[ \frac{\partial \varphi }{\partial t}   
+  \nabla_{\u,\x} \cdot \biggl(\varphi  \Biggl[ \begin{matrix} \dot{\u} \\ \u \end{matrix} \Biggr] \biggr)  \biggr ] \dU \dV  
\\&=   \iiint\limits_{\Omega(t)}  \iiint\limits_{\D(\x,t)}  \biggl[ \frac{\partial \varphi }{\partial t}   
+  \nabla_{\u} \cdot (\varphi   \dot{\u} )
+  \nabla_{\x} \cdot (\varphi   \u )
 \biggr ] \dU \dV  
\end{split}
\label{eq:Re_tr_velvol-temp}
\raisetag{65pt}
\end{align}
For separable integrals, further simplification is possible. The alternative path \pathcirc{4}-\pathcirc{2} can also be written using the second part of Eq.\ \eqref{eq:dens_varphi_int_double}. From Eq.\ \eqref{eq:Re_tr_velvol-temp} we can extract the continuity equation and differential equation based on $\loc{\ins{\varphi}}$ (Appendix \ref{sect:ApxC}), respectively:
\begin{align}
\begin{split}
 0
&= \frac{\partial \zeta  }{\partial t}   
+  \nabla_{\u} \cdot (\zeta \dot{\u})
+  \nabla_{\x} \cdot (\zeta \u )
\end{split}
\label{eq:DE_tr_velvol-temp_cty}
\\
 \zeta \frac{d \loc{\ins{\varphi}} }{dt}
&= \frac{\partial (\zeta \loc{\ins{\varphi}}) }{\partial t}   
+  \nabla_{\u} \cdot (\zeta \loc{\ins{\varphi}}   \dot{\u} )
+  \nabla_{\x} \cdot (\zeta \loc{\ins{\varphi}}   \u )
\label{eq:DE_tr_velvol-temp}
\end{align}

The conservation laws derived from Eqs.\ \eqref{eq:Re_tr_velvol-temp} and \eqref{eq:DE_tr_velvol-temp} are listed with their SI units in Table \ref{tab:Re_tr_velvol-temp_table}. 
Again the integrals equate to $dQ/dt$, hence to the same rates of change as the volumetric form (Table \ref{tab:Re_tr_vol-temp-alpha_table}), but are here identified with the phase volume PV$=\D \times \Omega$. The left-hand sides of the differential equations are written in the form of Eq.\ \eqref{eq:DE_tr_velvol-temp}. The corresponding Liouville equations, obtained from $\varphi(\u,\x,t)=p(\u,\x | t)$, are also listed in Table \ref{tab:Re_tr_velvol-temp_table}.

\subsection{Velocimetric-Temporal Formulation (Density $\varphi$)} \label{sect:exD}

Now consider a different velocimetric-temporal formulation, defined as in case \eqref{sect:exB}  but using the generalized density $\psi=\varphi(\u,\x,t)$ based on $\zeta(\u,\x,t)$. This follows the partial integration path \pathcirc{3} in Fig.\ \ref{fig:rel_diag_gen_densities}. From relation Eq.\ \eqref{eq:dens_varphi_int_left} between $\varphi$ and $\alpha$:
\begin{align}
\begin{split}
d\iiint\limits_{\D(\x,t)}  \varphi \, \dU 
= d \alpha 
=  \nabla_{\x} \alpha  \cdot d\x +  \frac{\partial \alpha}{\partial t}  dt 
\end{split}
\label{eq:alpha-rel}
\end{align}
Examining the time derivative term using Eq.\ \eqref{eq:Re_tr_gen}:
\begin{align}
\begin{split}
\frac{\partial \alpha}{\partial t} 
&= \frac{\partial }{\partial t} \iiint\limits_{\D(\x,t)}  \varphi \, \dU 
=   \iiint\limits_{\D(\x,t)}  \frac{\partial \varphi}{\partial t}  \, \dU 
+  \oiint\limits_{\partial \D(\x,t)} \varphi \dot{\u} \cdot \n_B \, dB 
\\  
&=    \iiint\limits_{\D(\x,t)}  \biggl[ \frac{\partial \varphi }{\partial t}   
+  \nabla_{\u} \cdot ( \varphi  \dot{\u} )  \biggr ] \dU 
\end{split}
\label{eq:Re_tr_vel-temp-varphi}
\raisetag{28pt}
\end{align}
For the conserved quantities considered, it is not straightforward to extract a differential equation from Eq.\ \eqref{eq:Re_tr_vel-temp-varphi}. 
Instead, it is necessary to write a Lie differential equation \eqref{eq:DE_gen_Lie} in terms of the Lie derivative $\mathcal{L}_{\dot{\u} \comp t}^{(t)}$ with respect to the augmented local acceleration field $\dot{\u} \comp t:= [\begin{smallmatrix} \dot{\u} \\ 1 \end{smallmatrix}]$. The left-hand term can then be treated as a placeholder for sources, sinks or drivers of the conserved quantity in the volume form $\vol^3_{\u}$. The conservation laws derived from Eq.\ \eqref{eq:Re_tr_vel-temp-varphi} are listed in Table \ref{tab:Re_tr_vel-temp-varphi_table}.  

For this formulation, it is possible to derive two sets of temporal Liouville equations, by assigning the density $\varphi(\u,\x,t)$ respectively to $p(\u,\x|t)$ or $p(\u|\x,t)$. The former gives a Lie differential equation, while the latter reduces by normalization in Eq.\ \eqref{eq:pdf_norm_pusubxt} to a partial differential equation. Both sets are listed in Table \ref{tab:Re_tr_vel-temp-varphi_table}.

\subsection{Volumetric-Temporal Formulation (Density $\varphi$)} \label{sect:exE}

Now consider a different volumetric-temporal formulation, defined as in case \eqref{sect:exA} but using the generalized density $\psi=\varphi(\u,\x,t)$ based on $\zeta(\u,\x,t)$. This follows the partial integration path \pathcirc{4} in Fig.\ \ref{fig:rel_diag_gen_densities}.
From Eq.\ \eqref{eq:dens_varphi_int_right} between $\varphi$ and $\beta$:
\begin{align}
\begin{split}
d\iiint\limits_{\Omega(\u,t)}  \varphi \, \dV
= d \beta 
=  \nabla_{\u} \beta  \cdot d\u + \frac{\partial \beta}{\partial t}  dt 
\end{split}
\label{eq:beta-rel}
\end{align}
Examining the time derivative term using Eq.\ \eqref{eq:Re_tr_gen}:
\begin{align}
\begin{split}
\frac{\partial \beta}{\partial t} 
&= \frac{\partial }{\partial t} \iiint\limits_{\Omega(\u,t)}  \varphi \, \dV 
=   \iiint\limits_{\Omega(\u,t)}  \frac{\partial \varphi}{\partial t}  \, \dV 
+  \oiint\limits_{\partial \Omega(\u,t)} \varphi {\u} \cdot \n \, dA 
\\  
&=    \iiint\limits_{\Omega(\u,t)}  \biggl[ \frac{\partial \varphi }{\partial t}   
+  \nabla_{\x} \cdot ( \varphi  {\u} )  \biggr ] \dV 
\end{split}
\raisetag{28pt}
\label{eq:Re_tr_vol-temp-varphi}
\end{align}
We again extract Lie differential equations \eqref{eq:DE_gen_Lie}, here written in terms of the Lie derivative $\mathcal{L}_{\u \comp t}^{(t)}$ with respect to the augmented velocity field $\u \comp t:= [\begin{smallmatrix} \u \\ 1 \end{smallmatrix}]$. The conservation laws derived from Eq.\ \eqref{eq:Re_tr_vol-temp-varphi} are listed in Table \ref{tab:Re_tr_vol-temp-varphi_table}.  We also derive two sets of temporal Liouville equations, now based on $p(\u,\x|t)$ or $p(\x|\u,t)$, listed in Table \ref{tab:Re_tr_vol-temp-varphi_table}.

\subsection{Velocimetric-Spatial (Time-Independent) Formulation (Density $\varphi$)} \label{sect:exF} 

Now consider a time-independent velocimetric-spatial formulation of the Reynolds transport theorem, based on the velocity space $\D$ with Cartesian velocity coordinates $\X=\u$, position parameter vector $\C=\x$, velocity gradient tensor field $\V=(\G(\u,\x))^\top:=(\nabla_{\x} \u)^\top$ and generalized density $\psi=\varphi(\u,\x)$, defined from the phase space fluid density $\zeta(\u,\x)$. This follows integration path \pathcirc{3} in Fig.\ \ref{fig:rel_diag_gen_densities}, but is independent of time, representing a stationary flow system\footnote{This formulation is suitable for statistically stationary flow systems -- for example a turbulent flow at steady state -- by mapping the time dependence to a velocity dependence, thus $\varphi(\u,\x,t) \mapsto \varphi(\u,\x, t(\u))$.}.
For the spatial gradient term in Eq.\ \eqref{eq:alpha-rel}, from Eq.\ \eqref{eq:Re_tr_gen} and relation \eqref{eq:dens_varphi_int_left} between $\varphi$ and $\alpha$ (for brevity omitting the surface integral formulation) \cite{Niven_etal_2020, Niven_etal_2020b}:
\begin{align}
\begin{split}
\nabla_{\x} \alpha
&= \nabla_{\x}  \iiint\limits_{\D(\x)} \varphi \, \dU
=   \iiint\limits_{\D(\x)} \biggl[  \nabla_{\x} \varphi   +  \nabla_{\u} \cdot (\varphi \G^\top) \biggr ] \dU 
\end{split}
\label{eq:Re_tr_vel-spat}
\end{align}
This gives an integral equation for the spatial gradient of the volumetric density $\alpha$. We can also write a Lie differential equation \eqref{eq:DE_gen_Lie}, expressed in terms of the multivariate Lie derivative $\mathcal{L}_{\G^\top \comp \x}^{(\x)}$ with respect to the augmented field $\G \comp \x:=\nabla_{\x} [\u,\x]$ over parameters $\x$.  The conservation laws derived from Eq.\ \eqref{eq:Re_tr_vel-spat} are listed in Table \ref{tab:Re_tr_vel-spat_table}.  We also present two sets of spatial Liouville equations based on $p(\u,\x)$ or $p(\u|\x)$ \cite{Niven_etal_2020}.

\subsection{Volumetric-Velocital (Time-Independent) Formulation (Density $\varphi$) }  \label{sect:exG}

Now consider a time-independent volumetric-velocital\footnote{We adopt the descriptor {\it velocital} from the Latin roots {\it velocitas} and {\it -al}, for ``pertaining to velocity''; in this usage we are preceded by Truesdell \cite{Truesdell_1954}.} formulation of the Reynolds transport theorem, based on the geometric space $\Omega$ with Cartesian coordinates $\X=\x$, velocity parameter vector $\C=\u$, inverse velocity gradient tensor field $\V=(\vGamma(\u,\x))^\top:=(\nabla_{\u} \x)^\top$ and generalized density $\psi=\varphi(\u,\x)$ defined from $\zeta(\u,\x)$. 
This formulation follows integration path \pathcirc{4} in Fig.\ \ref{fig:rel_diag_gen_densities}; recall that this adopts the velocimetric representation, which is integrated over the volumetric space for a distinct velocity. 
We again consider the time-independent case. From Eqs.\ \eqref{eq:beta-rel}, \eqref{eq:Re_tr_gen} and relation \eqref{eq:dens_varphi_int_right} between $\varphi$ and $\beta$:
\begin{align}
\begin{split}
\nabla_{\u} \beta
&= \nabla_{\u}  \iiint\limits_{\Omega(\u)} \varphi \, \dV
=   \iiint\limits_{\Omega(\u)} \biggl[ \nabla_{\u} \varphi   +  \nabla_{\x} \cdot (\varphi \vGamma^\top) \biggr ] \dV
\end{split}
\label{eq:Re_tr_vol-vel}
\end{align}
This gives the gradient in velocity space of the velocimetric density $\beta$. 
The conservation laws derived from Eq.\ \eqref{eq:Re_tr_vol-vel} are listed in Table \ref{tab:Re_tr_vol-vel_table}. 
We also list two sets of spatial Liouville equations based on $p(\u,\x)$ or $p(\x|\u)$.

\subsection{Velocimetric-Spatiotemporal Formulation (Density $\varphi$)}  \label{sect:exH}

Now consider the complete velocimetric-spatiotemporal formulation of the Reynolds transport theorem, a time-dependent extension of case \eqref{sect:exF} (path \pathcirc{3} in Fig.\ \ref{fig:rel_diag_gen_densities}). This examines the velocity space $\D$ with $\X=\u$, $\C=[\x,t]^\top$, $\V=(\wG(\u,\x,t))^\top:=(\delf_{\x} \u)^\top$, where $\delf_{\x} = \nabla_{\x,t}$ is the spatiotemporal gradient, and density $\psi=\varphi(\u,\x,t)$ defined from $\zeta(\u,\x,t)$.
Here $\wG = [\G, \dot{\u}]^\top$ is recognized as the velocity gradient tensor field $\G$ augmented with the local acceleration vector field $\dot{\u} := \partial \u/\partial t$. From Eqs.\ \eqref{eq:Re_tr_gen} and \eqref{eq:dens_varphi_int_left} \cite{Niven_etal_2020}:
\begin{align}
\begin{split}
\delf_{\x} \alpha
&= \delf_{\x}  \iiint\limits_{\D(\x,t)} \varphi \, \dU
=   \iiint\limits_{\D(\x,t)} \biggl[ \delf_{\x} \varphi   +  \nabla_{\u} \cdot (\varphi \wG^\top) \biggr ] \dU 
\end{split}
\label{eq:Re_tr_vel-spattemp}
\end{align}
This gives the spatiotemporal gradient $\delf_{\x} \alpha=[\nabla_{\x} \alpha, \dot{\alpha}]^\top$ of the volumetric density $\alpha$. 
The conservation laws derived from Eq.\ \eqref{eq:Re_tr_vel-spattemp} are listed in Table \ref{tab:Re_tr_vel-spattemp_table}.  We also present two sets of composite Liouville equations based on $p(\u,\x|t)$ or $p(\u|\x,t)$.

\subsection{Volumetric-Velocitemporal Formulation (Density $\varphi$) }  \label{sect:exI}

Now consider the complete volumetric-velocitemporal formulation of the Reynolds transport theorem, a time-dependent extension of that in \S\ref{sect:exG} (path \pathcirc{4} in Fig.\ \ref{fig:rel_diag_gen_densities}). This examines the position space $\Omega$ with $\X=\x$, $\C=[\u,t]^\top$, $\V=(\wGamma(\u,\x,t))^\top:=(\delf_{\u} \x)^\top$, where $\delf_{\u}=\nabla_{\u,t}$ is a velocitemporal gradient operator, and density $\psi=\varphi(\u,\x,t)$ defined from $\zeta(\u,\x,t)$. We recognize $\wGamma = [\vGamma,\u]^\top$ as the inverse velocity gradient tensor field $\vGamma$ augmented with the local velocity vector field $\u := \partial \x/\partial t$. From Eqs.\ \eqref{eq:Re_tr_gen} and \eqref{eq:dens_varphi_int_right}:
\begin{align}
\begin{split}
\delf_{\u} \beta
&=\delf_{\u}  \iiint\limits_{\Omega(\u,t)} \varphi \, \dV
=   \iiint\limits_{\Omega(\u,t)} \biggl[ \delf_{\u} \varphi  +  \nabla_{\x} \cdot (\varphi \widetilde{\vGamma}^\top) \biggr ] \dV
\end{split}
\label{eq:Re_tr_vol-veltemp}
\end{align}
This provides the velocitemporal gradient $\delf_{\u} \beta =[\nabla_{\u} \beta , \dot{\beta}]^\top$ of the velocimetric density $\beta$.  
The conservation laws derived from Eq.\ \eqref{eq:Re_tr_vol-veltemp} are listed in Table \ref{tab:Re_tr_vol-veltemp_table}.   We also list two sets of composite Liouville equations, here based on $p(\u,\x|t)$ or $p(\x|\u,t)$.

\subsection{Velocimetric-Temporal Formulation (Density $\theta$) } \label{sect:exJ} 
Now consider the alternative velocimetric-temporal formulation on path \pathcirc{5} in Fig.\ \ref{fig:rel_diag_gen_densities}, based on the generalized density $\psi=\theta(\u,\x,t)$ defined from the fluid density $\eta(\u,\x,t)$. 
From the definition of $\theta$ in Eqs.\ \eqref{eq:dens_theta_def} and \eqref{eq:dens_theta_norm}:
\begin{align}
\begin{split}
d  \iiint\limits_{\D(\x,t)} \theta \, \dU
= d Q(t)
=   \frac{dQ}{dt}  dt 
\end{split}
\label{eq:Q_theta_rel}
\end{align}
Note the peculiar property that integrating $\theta(\u,\x,t)$ with respect to $\u$ also eliminates $\x$, giving a velocimetric-temporal formulation with $\X=\u$, $\C=t$ and $\V=\dot{\u}$.
From Eq.\ \eqref{eq:Re_tr_gen}:
\begin{align}
\begin{split}
\frac{dQ}{dt} 
= \frac{d}{dt} \iiint\limits_{\D(\x,t)} \theta \, \dU 
&=   \iiint\limits_{\D(\x,t)}  \biggl[ \frac{\partial \theta }{\partial t}   +  \nabla_{\u} \cdot (\theta  \dot{\u}) \biggr ] \dU 
\end{split}
\label{eq:Re_tr_vel-temp-theta}
\raisetag{40pt}
\end{align}
The left-hand terms are again equivalent to those in Tables \ref{tab:Re_tr_vol-temp-alpha_table}-\ref{tab:Re_tr_velvol-temp_table}. 
The extracted continuity equation and differential equation based on $\ins{\loc{\theta}}$ (Appendix \ref{sect:ApxC}) are:
\begin{align}
\begin{split}
 0
&= \frac{\partial \eta }{\partial t}   +  \nabla_{\u} \cdot (\eta \dot{\u}) 
\end{split}
\label{eq:DE_tr_vel-temp-theta_cty}
\\
\begin{split}
 \eta \frac{d \ins{\loc{\theta}}}{dt}
&= \frac{\partial (\eta \ins{\loc{\theta}}) }{\partial t}   +  \nabla_{\u} \cdot (\eta \ins{\loc{\theta}} \dot{\u} ) 
\end{split}
\label{eq:DE_tr_vel-temp-theta}
\end{align}
The conservation laws derived from Eqs.\ \eqref{eq:Re_tr_vel-temp-theta} and \eqref{eq:DE_tr_vel-temp-theta} are listed  in Table \ref{tab:Re_tr_vel-temp-theta_table}. Apart from the different fluid density $\eta$ and specific density labels, these are mathematically identical to the conservation laws based on the velocimetric density $\de$ in Table \ref{tab:Re_tr_vel-temp-beta_table}. 
 The Liouville equations based on equating $\theta$ to its underlying pdf $p(\u|\x,t)$ are also listed, giving an alternative route to these equations to that given in Table \ref{tab:Re_tr_vel-temp-varphi_table}.

\subsection{Volumetric-Temporal Formulation (Density $\epsilon$)} \label{sect:exK}  
Now consider the alternative volumetric-temporal formulation on path $\pathcirc{6}$ in Fig.\ \ref{fig:rel_diag_gen_densities}, based on the generalized density $\psi=\epsilon(\u,\x,t)$ defined from the fluid density $\xi(\u,\x,t)$. From the definition of $\epsilon$ in Eqs.\ \eqref{eq:dens_epsilon_def} and \eqref{eq:dens_epsilon_norm}:
\begin{align}
\begin{split}
d  \iiint\limits_{\Omega(\u,t)} \epsilon \, \dV
= d Q(t)
=   \frac{dQ}{dt}  dt 
\end{split}
\label{eq:Q_epsilon_rel}
\end{align}
This also has the peculiar property that integration with respect to $\x$ eliminates $\u$, reducing this to a volumetric-temporal formulation with $\X=\x$, $\C=t$ and $\V={\u}$.  
From Eq.\ \eqref{eq:Re_tr_gen}:
\begin{align}
\begin{split}
\frac{dQ}{dt} 
= \frac{d}{dt} \iiint\limits_{\Omega(\u,t)} \epsilon \, \dV
&=   \iiint\limits_{\Omega(\u,t)}  \biggl[ \frac{\partial \epsilon }{\partial t}   +  \nabla_{\x} \cdot (\epsilon {\u}) \biggr ] \dV 
\end{split}
\label{eq:Re_tr_vol-temp-epsilon}
\raisetag{40pt}
\end{align}
with the left-hand terms again equivalent to those in Tables \ref{tab:Re_tr_vol-temp-alpha_table}-\ref{tab:Re_tr_velvol-temp_table}. 
The extracted continuity equation and differential equation based on $\ins{\loc{\epsilon}}$ (Appendix \ref{sect:ApxC}) are:
\begin{align}
\begin{split}
 0
&= \frac{\partial \xi }{\partial t}   +  \nabla_{\x} \cdot (\xi {\u}) 
\end{split}
\label{eq:DE_tr_vol-temp-epsilon_cty}
\\
\begin{split}
 \xi \frac{d \ins{\loc{\epsilon}}}{dt}
&= \frac{\partial (\xi \ins{\loc{\epsilon}}) }{\partial t}   +  \nabla_{\x} \cdot (\xi \ins{\loc{\epsilon}} {\u} ) 
\end{split}
\label{eq:DE_tr_vol-temp-epsilon}
\end{align}
The conservation laws derived from Eqs.\ \eqref{eq:Re_tr_vol-temp-epsilon} and \eqref{eq:DE_tr_vol-temp-epsilon} are listed  in Table \ref{tab:Re_tr_vol-temp-epsilon_table}. Apart from the fluid density $\xi$ and specific density labels, these are identical to the volumetric conservation laws based on $\rho$ in Table \ref{tab:Re_tr_vol-temp-alpha_table}. The Liouville equations based on the underlying pdf $p(\x|\u,t)$ are also listed, providing an alternative route to that in Table \ref{tab:Re_tr_vol-temp-varphi_table}.

\subsection{Summary} 

Summarizing the above analyses, we see that the first three and last two formulations (\eqref{sect:exA}-\eqref{sect:exC} and \eqref{sect:exJ}-\eqref{sect:exK}) are connected by their equivalence to the rate of change of the conserved quantity $dQ/dt$, and thence to the source-sink terms of the standard (volumetric) integral equations of fluid mechanics. These are based on complete integrations respectively of the densities $\alpha$, $\beta$, $\varphi$, $\theta$ and $\epsilon$. 

The remaining examples involve partial integrations of the density $\varphi$. Of these, the velocimetric-spatiotemporal formulation \eqref{sect:exH} combines the velocimetric-temporal \eqref{sect:exD} and velocimetric-spatial \eqref{sect:exF} formulations connected by Eq.\ \eqref{eq:alpha-rel}, based on $\alpha$ and $\varphi$. Similarly, the volumetric-velocitemporal formulation \eqref{sect:exI} combines the volumetric-temporal \eqref{sect:exE} and volumetric-velocital \eqref{sect:exG} formulations connected by Eq.\ \eqref{eq:beta-rel}, based on $\beta$ and $\varphi$. For these formulations, it is possible to extract a Lie differential equation based on a Lie operator defined for that formulation. 

All formulations can be used to derive a number of Liouville equations based on the underlying pdfs defined in \S \ref{sect:pdfs} and Fig.\ \ref{fig:rel_diag_pdfs}.  In general, these are in the form of Lie differential equations, but for cases in which the pdf is normalized by the integration, the analysis yields a homogeneous integral equation and associated partial differential equation.

\section{\label{sect:concl} Conclusions}

This study examines recent generalizations of the Reynolds transport theorem \cite{Niven_etal_2020, Niven_etal_2020b}, which provide continuous multiparametric mappings of a differential form on a manifold -- or of a density within a generalized coordinate space -- connected by the maximal integral curves described by a vector or tensor field. 
These extend the formulation of Reynolds \cite{Reynolds_1903} to encompass new {\it transformation theorems}, which exploit previously unreported multiparametric continuous (Lie) symmetries associated with a conserved quantity in the space considered. In the present study, we explore the implications of this generalized framework for fluid flow systems, within an extended Eulerian position-velocity (phase space) description instead of the traditional Eulerian volumetric description.

The analysis starts from a hierarchy of the five pdfs $p(\x|t), p(\u|t), p(\u,\x|t), p(\u|\x,t)$ and $p(\x|\u,t)$, examined in \S\ref{sect:pdfs}. In \S\ref{sect:fluid_densities}-\ref{sect:gen_densities}, these are used to define analogous hierarchies of fluid densities $\rho$, $\de$, $\zeta$, $\eta$ and $\xi$ and generalized densities $\alpha$, $\beta$, $\varphi$, $\theta$ and $\epsilon$, applicable to different spaces. The generalized Reynolds transport theorems are then examined in \S\ref{sect:ReTT}, in both exterior calculus and vector calculus formulations. Their connections to underlying partial differential equations and Lie differential equations -- the latter containing the Lie derivative of a volume form in the domain -- are also examined. In \S\ref{sect:ex_sys}, the densities and theorems are used to obtain 11 formulations of the Reynolds transport theorem arising from the position-velocity description, for different choices of the coordinate space, parameter space and density. 
These are reported in the form of 11 tables of integral and differential conservation laws applicable to these systems, for the eight common conserved quantities of interest in fluid dynamics (fluid mass, species mass, linear momentum, angular momentum, energy, charge, entropy and probability). 
The integral equations include expressions for the time derivative of the total conserved quantity $dQ/dt$, for the time derivative or spatial gradient of the volumetric density $\alpha$, or for the time derivative or velocity gradient of the velocimetric density $\beta$, expressed in terms of lower order densities within the hierarchy. 
These provide a considerable assortment of new conservation laws for the analysis of fluid flow systems.


While every effort has been made to provide a comprehensive treatment, this study exclusively considers Cartesian position and velocity coordinates, with no attempt to examine orthogonal or non-orthogonal curvilinear coordinate systems or parameter spaces (however, care is taken to distinguish the vector of partial derivatives $\vpartial_{\X}$ from the gradient $\nabla_{\X}$). Further work is required to extend these analyses to general coordinate systems.  Further detailed study is also required of the partial and Lie differential equations reported in Tables \ref{tab:Re_tr_vol-temp-alpha_table}-\ref{tab:Re_tr_vol-temp-epsilon_table}, to identify the appropriate source-sink terms for each expression. The analyses are also restricted to fixed frames of reference (control volumes or domains), but can readily be extended to moving and deforming frames of reference using relative vector or tensor gradients $\V_{rel}$ \citep[e.g.,][]{White_1986, Munson_etal_2010, Niven_etal_2020}. The analyses could also be united with insights from other studies, to consider domains with jump discontinuities \cite{Truesdell_Toupin_1960, Dvorkin_Goldschmidt_2005}, or irregular and fragmenting domains \cite{Seguin_Fried_2014, Seguin_etal_2014, Falach_Segev_2014}.  The generalized Reynolds theorem framework can also be used to generate conservation laws for many other dynamical systems containing conserved quantities \citep[c.f.,][]{Niven_etal_2020}.

\begin{acknowledgments}
This work was largely completed during the COVID-19 lockdown of Canberra, Australia, in March-May 2020, based on research supported by UNSW and Institute Pprime, CNRS, Poitiers, France. Michael Schlegel is thanked for comments on the manuscript.
\end{acknowledgments}

\section*{Data Availability}
Data sharing is not applicable to this article as no new data were created or analyzed in this study.

\appendix

\section{\label{sect:ApxA}Definitions of Densities}

The definition of a density based on spatial averaging has received considerable attention in the literature, especially for the analysis of fluid flow through porous media, where many different averages can be defined and a precise terminology is necessary 
\citep[e.g.,][]{Matheron_1965, Marle_1967, Anderson_Jackson_1967, Whitaker_1967a, Whitaker_1969, Bachmat_1972, Gray_Lee_1977, Hassanizadeh_Gray_1979a, Narasimhan_1980, Ene_1981, Cushman_1982, Cushman_1983b, Baveye_Sposito_1984, Cushman_1984, Cushman_1985, Bear_Bachmat_1991, Gray_etal_1993, Quintard_Whitaker_1993, Chen_1994, Grau_Cantero_1994, Cushman_1997, Gray_Miller_2005b, Whitaker_1999, Wood_2013, Gray_Miller_2013, Pokrajac_deLemos_2015, Davit_Quintard_2017, Takatsu_2017}.  Consider the generalized volumetric density $\alpha(\x,t)$ [qty m\sup{-3}] in Eq.\ \eqref{eq:a_def} based on the fluid density $\rho(\x,t)$, which many authors define by the convolution
\cite{Matheron_1965, Whitaker_1967a, Whitaker_1969, Bachmat_1972, Gray_Lee_1977, Cushman_1982, Hassanizadeh_Gray_1979a, Bear_Bachmat_1991, Gray_etal_1993, Quintard_Whitaker_1993, Grau_Cantero_1994, Cushman_1997, Whitaker_1999, Wood_2013, Pokrajac_deLemos_2015, Davit_Quintard_2017, Takatsu_2017}:
\begin{align}
\lexp \alpha\rexp(\x,t)  = \frac{1}{V(\x,t)} \iiint\nolimits_{\Vsp(\x,t)}  \alpha(\x + \r, t) \, \dV(\r, t)
\label{eq:a_def_volume}
\end{align}
where $\lexp \cdot \rexp$ indicates a volumetric expected value, $\Vsp$ is a small fluid domain, $V = | \Vsp |$ [m\sup{3}] is the volume of $\Vsp$, and here $\dV=dr_x dr_y dr_z = d^3 \r$ is an infinitesimal volume element in $\Vsp$. 
Usually Eq.\ \eqref{eq:a_def_volume} has a geometric interpretation, but $V$ can be reinterpreted as the measure of set $\Vsp$ \cite{Cushman_1983b, Chen_1994, Cushman_1997}. 
As noted, most authors require $\Vsp$ to be sufficiently large that the medium can be considered a continuum, but smaller than any larger-scale heterogeneities.
However, some authors reject the notion of an averaging volume as a continuum property, to instead recognize it as a characteristic of the measurement scale \cite{Baveye_Sposito_1984, Cushman_1984, Cushman_1985, Chen_1994, Cushman_1997}.
In multiphase systems, variants of Eq.\ \eqref{eq:a_def_volume} can be defined for each phase, in which many authors use a  characteristic function $\theta_\alpha$ to indicate the presence of phase $\alpha$
\cite{Gray_Lee_1977, Hassanizadeh_Gray_1979a, Cushman_1982, Cushman_1983b, Baveye_Sposito_1984, Cushman_1984, Bear_Bachmat_1991, Gray_etal_1993, Quintard_Whitaker_1993, Chen_1994, Grau_Cantero_1994, Pokrajac_deLemos_2015}.  
Eq.\ \eqref{eq:a_def_volume} has also been generalized to give a variety of averages in \mbox{zero-}, \mbox{one-}, \mbox{two-} and three-dimensional spaces and/or time 
\cite{Whitaker_1969, Bachmat_1972, Hassanizadeh_Gray_1979a, Bear_Bachmat_1991, Gray_etal_1993, Grau_Cantero_1994, Whitaker_1999, Gray_Miller_2005b, Gray_Miller_2013}, 
and used to examine the effects of multiple or time-varying averaging scales 
\cite{Hassanizadeh_Gray_1979a, Cushman_1984, Gray_etal_1993, Quintard_Whitaker_1993, Chen_1994, Cushman_1997, Whitaker_1999, Gray_Miller_2005b, Gray_Miller_2013, Wood_2013, Pokrajac_deLemos_2015, Takatsu_2017}.

Eq.\ \eqref{eq:a_def_volume} can be further generalized using a weighting function
\cite{Monin_Yaglom_1971a, Matheron_1965, Marle_1967, Anderson_Jackson_1967, Ene_1981, Baveye_Sposito_1984, Cushman_1984, Quintard_Whitaker_1993, Chen_1994, Cushman_1997, Davit_Quintard_2017}, here labelled $w$:
\begin{align}
\lexp \alpha\rexp(\x,t)  =  \iiint\nolimits_{\Vsp(\x,t)}  w(\r, t) \, \alpha(\x + \r, t) \, \dV(\r, t),
\label{eq:a_def_weight}
\end{align}
The weighting function enables each volume element to be weighted differently, and indeed is chosen by most adherents to allow integration over the entire domain $\Omega(t)$. 
If assumed uniformly distributed over the small averaging volume $w=V^{-1}$, Eq.\ \eqref{eq:a_def_weight} reduces to  Eq.\ \eqref{eq:a_def_volume}. 
As commonly defined \cite{Anderson_Jackson_1967, Baveye_Sposito_1984}, $w$ is infinitely differentiable, $w \ge 0$ and $\iiint\nolimits_{\Omega(t)}  w(\r, t)  \, \dV(\r, t)=1$, from which $w$ has units of reciprocal volume, and can be seen to satisfy the properties of the conditional pdf $p(\r | t)$. 

Other definitions of continuum averages have also been proposed. Many authors \citep[e.g.,][]{Monin_Yaglom_1971a, Marle_1967, Ene_1981, Cushman_1982, Grau_Cantero_1994} have extended Eq.\ \eqref{eq:a_def_volume} or \eqref{eq:a_def_weight} by integration over time to give temporal and spatiotemporal averages. 
Cushman \cite{Cushman_1982, Cushman_1983b, Cushman_1985} defined a density by the mathematical limit of a distribution, and show it reduces to Eq.\ \eqref{eq:a_def_volume} or \eqref{eq:a_def_weight} under certain assumptions.  Cushman \cite{Cushman_1984, Cushman_1985, Cushman_1997} also identified the weighted average Eq.\ \eqref{eq:a_def_weight} as a frequency filter (later exploited for periodic media \cite{Davit_Quintard_2017}) and as a compactness filter, in which $w$ plays the role of a test function (see also \cite{Quintard_Whitaker_1993}) and need not satisfy a differentiability property.

Several authors \cite{Hassanizadeh_Gray_1979a, Narasimhan_1980, Ene_1981} also reemphasized that the volume average Eq.\ \eqref{eq:a_def_volume} applies only to extensive variables, and intensive variables such as temperature or pressure require additional ``capacity'' terms (thermodynamic susceptibilities) in the integrand and divisor \cite{Narasimhan_1980}, or should be incorporated without averaging \cite{Hassanizadeh_Gray_1979a}. This concept was extended in a series of many papers by Gray, Miller and co-authors on thermodynamically constrained averaging theory \citep[e.g.,][]{Gray_Miller_2005b, Gray_Miller_2013} based on unnormalized geometric or  thermodynamic weighting functions, such that the right-hand side of Eq.\ \eqref{eq:a_def_weight} is normalized by $\iiint\nolimits_{\Vsp(\x,t)}  w(\r, t)  \, \dV(\r, t)$ (or its equivalent in discrete spaces). 


In the present study, we formally identify the weighting function as a pdf (as foreshadowed in \cite{Marle_1967, Cushman_1983b, Cushman_1984}), to define the expected generalized volumetric density in terms of its underlying pdf:
\begin{align}
\lexp \alpha\rexp(\x,t)  =  \iiint\nolimits_{\Vsp(\x,t)}  p(\r | t) \, \alpha(\x + \r, t) \, \dV(\r, t)
\label{eq:a_def_prob_vol}
\end{align}
This contains the density $\alpha(\x + \r, t)$ at each point, which can be reduced by the differential relation:
\begin{align}
dQ (\x+\r,t)= \alpha(\x + \r, t) \, \dV (\r,t)
\label{eq:a_point_def}
\end{align}
where $dQ$ [qty] is a differential element of the conserved quantity.  Coordinate transformation of Eq.\ \eqref{eq:a_def_prob_vol} then gives:
\begin{align}
\lexp \alpha\rexp(\x,t)  =  \int\nolimits_{\Qsp(\x,t)}  p(\r | t) \,  dQ(\x+\r, t)
\label{eq:a_def_prob_qty}
\end{align}
where $\Qsp$ is the small quantity domain corresponding to $\Vsp$. Using the relation in Eq.\ \eqref{eq:a_def} this also gives:
\begin{align}
\lexp \alpha\rexp(\x,t)  =  \int\nolimits_{\msp(\x,t)}  p(\r | t) \, \loc{\alpha}(\x+r, t) \, dm(\x+\r, t)
\label{eq:a_def_prob_qty2}
\end{align}
based on the specific density $\loc{\alpha}$ of the conserved quantity.

As an example, consider the application of Eq.\ \eqref{eq:a_def_prob_vol} to the volumetric fluid density $\rho$, giving: 
\begin{align}
\lexp \rho\rexp (\x,t) 
&= \iiint\nolimits_{\Vsp(\x,t)} p(\r|t) \, \rho(\x + \r, t) \, \dV(\r,t) 
\label{eq:rho_def_prob_vol}
\end{align}
Using the differential relation:
\begin{align}
dm (\x+\r,t)= \rho(\x + \r, t) \, \dV (\r,t) 
\label{eq:rho_point_def}
\end{align}
this reduces to: 
\begin{align}
\lexp \rho\rexp (\x,t) 
&= \int\nolimits_{\msp(\x,t)} p(\r|t) \, dm(\x + \r, t)  
\label{eq:rho_def_prob_mass}
\end{align}
thus giving both parts of Eq.\ \eqref{eq:pxsubt_rho}.

Since the volume integral definitions in Eqs.\ \eqref{eq:a_def_prob_vol} and \eqref{eq:rho_def_prob_vol} require knowledge of the density at each point, the mass and quantity integrals in Eqs.\ \eqref{eq:rho_def_prob_mass} and \eqref{eq:a_def_prob_qty} -- and their analogs in different spaces -- are adopted as the primary definitions respectively of the fluid and generalized densities in this study. 

We note that some authors \citep[e.g.,][]{Spurk_1997} consider each element in the mass integrals such as Eq.\ \eqref{eq:rho_def_prob_mass} -- and by implication in the quantity integrals such as Eq.\ \eqref{eq:a_def_prob_qty} -- to be represented by their original (Lagrangian) position coordinate $\x_0$, for each position and time. This would require a coordinate transformation between the local and Lagrangian coordinates, e.g., for Eq.\ \eqref{eq:rho_def_prob_mass}:
\begin{align}
\begin{split}
&\lexp \rho\rexp (\x,t) \\
&= \int\nolimits_{\msp(\x_0)} p(\r(\x_0)|t(\x_0)) \, 
\biggl | \frac{\partial \r}{\partial \x_0} \biggr |
dm( (\x + \r)(\x_0), t(\x_0))  
\end{split}
\label{eq:rho_def_prob_mass_Lagr}
\end{align}
where $ | {\partial \r}/{\partial \x_0} |$ is the determinant of the coordinate Jacobian. 
This construction is correct, but could create many complications, for example if the Lagrangian mass domain $\msp(\x_0)$ consists of disjoint regions, creating singularities in the Jacobian. It is for this reason that we consider the mass domain $\msp(\x,t)$ or quantity domain $\Qsp(\x,t)$ to be in one-to-one correspondence (bijective) with -- and to have the same topology as -- the local volume domain $\Vsp(\x,t)$, based on differential relations such as Eqs.\ \eqref{eq:a_point_def} or \eqref{eq:rho_point_def}.

As defined here, the fluid and generalized densities are nonnegative, corresponding to their underlying pdfs. However, their definitions extend naturally to negative conserved quantities, for example systems containing matter and antimatter, or positive and negative charged particles, by considering a fluid mass or conserved quantity domain defined in $\R$ rather than $\R^+_0$, leading to negative densities. Even more generalized formulations, which allow for positive and negative volumes and local coordinate systems, are available using differential forms such as $\rho dx \wedge dy \wedge dz$ in the framework of exterior calculus \cite{Niven_etal_2020}.

\section{\label{sect:ApxB}Philosophical Implications}

In probability theory, there are two main philosophical interpretations of probabilities:
\begin{enumerate}
\item The {\it frequentist interpretation}, in which probabilities are considered to represent measurable frequencies. In this viewpoint, a probability distribution is equivalent to the frequency distribution of an infinite number of random samples collected from a stationary sample space \citep[e.g.,][]{Feller_1966}.

\item The {\it Bayesian} or {\it probabilistic interpretation}, in which a probability is a mathematical assignment based on one's knowledge, which need not correspond to a measurable frequency. Nonetheless, a probability is a rational assignment, which can be calculated and manipulated using the rules of probability theory \cite{Cox_1961, Jaynes_2003}. 
\end{enumerate}

In continuum fluid mechanics, it is generally considered that every continuum variable, such as the density $\rho(\x,t)$ or velocity $\u(\x,t)$, will have a ``true'' value at each position and time in a given flow system. Each such variable is considered to be in principle measurable, regardless 
of the difficulties inherent in its measurement. In this viewpoint, the definitions of the fluid densities in terms of their underlying pdfs in Eqs.\ \eqref{eq:pxsubt_rho}-\eqref{eq:puxsubt_xi} suggest that these pdfs are also measurable -- in accordance with the frequentist interpretation -- and indeed, these relations provide a tool for their measurement.

The definitions in Eqs.\ \eqref{eq:pxsubt_rho}-\eqref{eq:puxsubt_xi} do however open the door to a very different {\it statistical} or {\it probabilistic} interpretation of continuum variables as assignments based on one's knowledge of the flow system. In practice, it can be difficult to accurately measure the density of a compressible fluid at a given position and time, while the measurement of the three-dimensional velocity in a turbulent flow system will always require some degree of time and spatial averaging and interference with the flow.  Furthermore, in a turbulent flow, there will always be a stochastic or probabilistic component of each continuum variable, which cannot be predicted accurately by deterministic methods. Indeed, an observer will usually have far less information than that needed to conduct such a deterministic analysis. The definitions in Eqs.\ \eqref{eq:pxsubt_rho}-\eqref{eq:puxsubt_xi} therefore suggest that continuum variables can be interpreted as probabilistic quantities, which must be inferred from the limited data available using the laws of physics (especially the conservation equations) and the rules of probability theory.

\section{\label{sect:ApxC}Extraction of Differential Equations}

For several integral equations, we can extract the differential equation by a simple manipulation \cite{Hutter_Johnk_2004}. 
First considering Eq.\ \eqref{eq:Re_tr_vol-temp-alpha}, we substitute $\alpha = \rho \loc{\alpha}$ in Eq.\ \eqref{eq:a_def} based on the local specific density $\loc{\alpha}$, to give:
\begin{align}
\begin{split}
&\frac{d}{dt}  \iiint\limits_{\Omega(t)} \rho \loc{\alpha} \, \dV 
=   \iiint\limits_{\Omega(t)}  \biggl[ \frac{\partial (\rho \loc{\alpha}) }{\partial t}   +  \nabla_{\x} \cdot (\rho \loc{\alpha}  \u ) \biggr ] \dV  
\\
&= \iiint\limits_{\Omega(t)} \biggl\{ \rho \biggl[ \frac{\partial \loc{\alpha}}{\partial t} + \u \cdot \nabla_{\x} \loc{\alpha} \biggr] 
+ \loc{\alpha} \biggl [ \frac{\partial \rho}{\partial t} + \nabla_{\x} \cdot (\rho \u) \biggr] \biggr\} \dV
\end{split}
\label{eq:Re_tr_spec}
\raisetag{50pt}
\end{align}
For the conservation of fluid mass, $\loc{\alpha}=1$ and  from Eq.\ \eqref{eq:dens_rho_norm} $\frac{d}{dt} \iiint\nolimits_{\Omega(t)} \rho \, \dV$ $=dM/dt=0$, giving:
\begin{align}
\begin{split}
0 &=  \iiint\limits_{\Omega(t)} \biggl [ \frac{\partial \rho}{\partial t} + \nabla_{\x} \cdot (\rho \u) \biggr] \dV
\end{split}
\label{eq:Re_tr_cty}
\end{align}
Invoking the fundamental lemma of the calculus of variations \cite{Weinstock_1952, Gelfand_Fomin_1963}, thus for a compactly supported, continuous and continuously differentiable density $\rho$, this gives the continuity equation:
\begin{align}
\begin{split}
 \frac{\partial \rho}{\partial t} + \nabla_{\x} \cdot (\rho \u) =0
\end{split}
\label{eq:DE_tr_cty}
\end{align}
Substitution into the last term in Eq.\ \eqref{eq:Re_tr_spec} gives:
\begin{align}
\begin{split}
\frac{d}{dt}  \iiint\limits_{\Omega(t)} \rho \loc{\alpha} \, \dV 
&= \iiint\limits_{\Omega(t)}  \rho \biggl[ \frac{\partial \loc{\alpha}}{\partial t} + \u \cdot \nabla_{\x} \loc{\alpha} \biggr] 
 \dV
\end{split}
\label{eq:Re_tr_spec2}
\end{align}
Equating Eqs.\ \eqref{eq:Re_tr_spec} and \eqref{eq:Re_tr_spec2}, and again invoking the fundamental lemma of the calculus of variations \cite{Weinstock_1952, Gelfand_Fomin_1963} -- thus for compactly supported, continuous and continuously differentiable $\rho$ and $\loc{\alpha}$ -- we obtain the differential equation \eqref{eq:DE_tr_vol-temp-alpha}:
\begin{align}
\begin{split}
\frac{\partial (\rho \loc{\alpha}) }{\partial t}   +  \nabla_{\x} \cdot (\rho \loc{\alpha} \u ) 
&= \rho \biggl[ \frac{\partial \loc{\alpha}}{\partial t} + \u \cdot \nabla_{\x} \loc{\alpha} \biggr] 
\\&= \rho \frac{D \loc{\alpha}}{Dt}
= \rho \frac{d \loc{\alpha}}{dt}
\end{split}
\label{eq:DE_tr_RHS2}
\end{align}
We recognize that the second term can be written in terms of the substantial derivative $D/Dt$, here equivalent to the total derivative $d/dt$. 

Similar manipulations of Eqs.\ \eqref{eq:Re_tr_vel-temp-beta}, \eqref{eq:Re_tr_velvol-temp}, \eqref{eq:Re_tr_vel-temp-theta} and \eqref{eq:Re_tr_vol-temp-epsilon} respectively yield 
the continuity equations \eqref{eq:DE_tr_vel-temp-beta_cty}, \eqref{eq:DE_tr_velvol-temp_cty}, \eqref{eq:DE_tr_vel-temp-theta_cty} and \eqref{eq:DE_tr_vol-temp-epsilon_cty}
and general differential equations \eqref{eq:DE_tr_vel-temp-beta}, \eqref{eq:DE_tr_velvol-temp}, \eqref{eq:DE_tr_vel-temp-theta} and \eqref{eq:DE_tr_vol-temp-epsilon}.

\section{\label{sect:ApxD}Nomenclature}
\setcounter{table}{11}    

The mathematical operators and symbols used in this study are listed in Table \ref{tab:nomen}.

\newpage
\begin{longtable}{p{90pt} p{360pt}}  
\caption{\label{tab:nomen}Nomenclature used in this study.} \\ 
\hline\hline
{\bf Symbol} &{\bf Description, [SI Units]}  \\
\hline 
 \endhead
\multicolumn{2}{l}{\bf Mathematical Operators} \\
$^{\top}$ &transpose \\
$\cdot$ &vector scalar product \\
$\times$ &cross product, or multiplication symbol (only if line break in equation) \\
$\partial$ &partial derivative operator, or boundary of domain \\
$\vpartial_{\C}$ &vector partial derivative operator with respect to $\C$ \\
$\nabla_{\C}$ &gradient operator with respect to $\C$ \\
$\nabla_{\u}$ &velocital gradient operator with respect to $\u$, [(m s\sup{-1})\sup{-1}] \\
$\nabla_{\x}$ &spatial gradient operator with respect to $\x$, [m\sup{-1}] \\  
$\nabla_{\X}$ &gradient operator with respect to $\X$ \\
$\delf_{\u}=\nabla_{\u,t}$ &velocitotemporal gradient operator with respect to $[\u,t]$, components [(m s\sup{-1})\sup{-1}] or [s\sup{-1}]\\
$\delf_{\x}=\nabla_{\x,t}$ &spatiotemporal gradient operator with respect to $[\x,t]$, components [m\sup{-1}] or [s\sup{-1}] \\
$\lexp \cdot \rexp$ &expectation over small volume $\Usp$ \\
$\langle \cdot \rangle$ &expectation over small velocity domain $\Vsp$ \\
$\lvol \cdot \rvol$ &integral over volume $\Omega$ \\
$\lvel \cdot \rvel$ &integral over velocity domain $\D$ \\
$\wedge$ &wedge product \\
$\comp$ &augmentation operator, such that $\V \comp \C$ is the tensor $\V$ based on coordinates $\X$ augmented by parameter $\C$ \\
$\pathcirc{1}$ &integral path as labelled in Fig. \ref{fig:rel_diag_gen_densities} \\
\hline
\multicolumn{2}{l}{\bf Conventions}\\
\multicolumn{2}{l}{Vector derivatives are defined by the $\partial (\to)/\partial (\downarrow)$  convention} \\
\multicolumn{2}{l}{The product of two vectors implies a tensor, e.g., $\u \u := \u \u^\top$} \\
\multicolumn{2}{l}{The divergence of a tensor is rotated, e.g., $\nabla_{\x} \cdot \G:= (\nabla_{\x}^\top \G)^\top$} \\
\hline
\multicolumn{2}{l}{\bf Roman symbols}\\
$c$ &index of chemical species, or index of components of $\C$ \\
$C_c$ &$c$th component of $\C$ \\
$\C$ &generalized $m$-dimensional parameter vector \\
$CV$ &control volume = reference frame for fluid motion \\
$d$ &differential of a function, or exterior derivative of a differential form \\
$\hat{d}$ &extended exterior derivative based on augmented coordinates \\
$d/dt$ &total derivative in time, [s\sup{-1}] \\
$dA$ &infinitesimal area element in volumetric space, [m\sup{2}] \\
$d\A$ &directed infinitesimal area element in volumetric space, [m\sup{2}] \\
$dB$ &infinitesimal surface element in velocimetric space, [{(m s\sup{-1})\sup{2}}] \\
$d\B$ &directed infinitesimal surface element in velocimetric space, [{(m s\sup{-1})\sup{2}}] \\
$dm$ &infinitesimal element of fluid mass, [kg] \\
$\dU=dudvdw$ &infinitesimal element of velocimetric space, [(m s\sup{-1})\sup{3}] \\
$\dV=dxdydz$ &infinitesimal element of volumetric space, [m\sup{3}] \\
$dX_j$ &cotangent to $j$th component of generalized vector $\X$ \\
$d^{n-1} \X$ &generalized directed area element on $\partial \Omega$ \\
$d^n \X$ &generalized volume element in $\Omega$ \\
$\vecunit_{\x,t}$ &vector of spatiotemporal SI units = [m,m,m,s]$^\top$ \\
$\vecunit_{\u,t}$ &vector of velocitotemporal SI units = [m s\sup{-1},m s\sup{-1},m s\sup{-1},s]$^\top$ \\
$D/Dt$ &substantial or material derivative in time, [s\sup{-1}] \\
$\D$ &velocity domain \\
$\loc{e}$ &local specific total energy, [J kg\sup{-1}] \\
$\ins{e}$ &velocity-distinct specific energy, [J kg\sup{-1}] \\
$\ins{\loc{e}}$ &velocity-distinct local specific energy, [J kg\sup{-1}] \\
$E$ &total energy, [J] \\
$\sum \vec{F}$ &sum of forces, [N] \\
$FV = \Omega$ &fluid volume \\
$\vec{g}$ &acceleration due to gravity, [m s\sup{-2}] \\
$\G := \nabla_{\x} \u$ &velocity gradient tensor field, [(m s\sup{-1}) m\sup{-1}] =  [s\sup{-1}] \\
$\wG := \delf_{\x} \u $ &augmented velocity gradient tensor field, components [(m s\sup{-1}) m\sup{-1}] or [(m s\sup{-1}) s\sup{-1}] 
\\
$\vec{i}$ &electrical flux, [C m\sup{-2} s\sup{-1}] = [A m\sup{-2}] \\
$i_{\V}^{(\C)}$ &multivariate interior product with respect to $\V$ over parameters $\C$ \\
$I$ &net inward electrical current (passive sign convention), [C s\sup{-1}] = [A]\\
$\I_m$ &identity matrix of size $m$ \\
$j$ &index of components of generalized coordinates $\X$ \\
$\vec{j}_c$ &molar flux of species $c$, [mol m\sup{-2} s\sup{-1}] \\
$\vec{j}_Q$ &heat flux, [J m\sup{-2} s\sup{-1}] \\
$\vec{j}_S$ &entropy flux, [J K\sup{-1} m\sup{-2} s\sup{-1}]  \\
$\mathcal{L}_{\V}^{(\C)}$ &multivariate Lie derivative with respect to $\V$ over parameters $\C$ \\
$\mathcal{L}_{\V \comp \C}^{(\C)}$ &multivariate Lie derivative with respect to $\V \comp \C$ over parameters $\C$ \\
LHS &left-hand side \\
$\msp$ &small fluid mass domain \\
$\dot{m}$ &rate of production of fluid mass, [kg s\sup{-1}] \\
$\dot{m}_c$ &rate of production of mass of species $c$, [kg\sub{c} s\sup{-1}] \\
$m$  &dimension of vector parameter $\C$ \\
$M$ &total fluid mass, [kg] \\ 
$\dot{M}$ &rate of change of total fluid mass, [kg s\sup{-1}] \\ 
$M_c$ &molar mass of species $c$, [kg\sub{c} mol\sup{-1}] \\
$\dot{M}_c$ &rate of change of mass of species $c$, [kg\sub{c} s\sup{-1}] \\ 
$\Msp$ &orientable differentiable manifold, or generalized space \\
$n$  &dimension of manifold $M$, dimension of coordinates $\X$ \\
$\n$ &outward unit normal to $\Omega$ in volumetric space \\
$\n_B$ &outward unit normal to $\D$ in velocimetric space \\
$p(a, b | c) = p_{a, b | c}$  &conditional pdf of $a$ and $b$ subject to $c$, [units of $(ab)^{-1}$] \\
$P$ &pressure, [Pa] = [J m\sup{-3}] \\
$PV = \D \times \Omega$ &phase volume \\
$Q$ &total conserved quantity (of any type), [qty] \\
$\dot{Q}_{in}$ &net inward heat flow rate, [J s\sup{-1}] \\
$r$ &dimension of submanifold $\Omega$, dimension of differential form $\omega^r$ \\
$\r = [r_x, r_y, r_z]^\top$ &local Cartesian position coordinates, [m] \\
$\loc{\r}$ &local radius of a lever arm, [m] \\
$\ins{\r}$ &velocity-distinct radius of a lever arm, [m] \\
$\loc{\ins{\r}}$ &velocity-distinct local radius of a lever arm, [m] \\
RHS &right-hand side \\
$\loc{s}$ &local specific entropy, [J K\sup{-1} kg\sup{-1}] \\
$\ins{s}$ &velocity-distinct specific entropy, [J K\sup{-1} kg\sup{-1}] \\
$\ins{\loc{s}}$ &velocity-distinct local specific entropy, [J K\sup{-1} kg\sup{-1}] \\
$\s= [s_u, s_v, s_w]^\top$ &local Cartesian velocity coordinates, [m s\sup{-1}] \\
$S$ &total entropy, [J K\sup{-1}] \\
$\dot{S}^{nf}$ &total net inward non-fluid entropy flow rate, [J K\sup{-1} s\sup{-1}] \\
$t$ &time, [s] \\
$\sum \vec{T}$ &sum of torques, [N m] \\
$\u = [u,v,w]^\top := \frac{\partial \x}{\partial t}$ &Cartesian velocity field, [m s\sup{-1}] \\
$\dot{\u} = [\dot{u},\dot{v},\dot{w}]^\top :=\frac{\partial \u}{\partial t}$ & Cartesian local acceleration field, [m s\sup{-2}] \\
$\Usp$ &small velocity domain \\
$\vol^n_{\X}$ &volume of an infinitesimal $n$-dimensional parallelopiped spanned by the cotangents to $\X$ \\
$V_{jc}$ &$(jc)$th component of generalized vector or tensor field $\V$ \\
$\V$ &generalized vector or tensor field \\
$\Vsp$ &small fluid volume \\
$VV = \D$ &velocity volume \\
$\dot{W}_{in}$ &net inward work flow rate, [J s\sup{-1}] \\
$\x = [x,y,z]^\top$ &Cartesian position coordinates, [m] \\
$\x_0= [x_0,y_0,z_0]^\top$ &Cartesian Lagrangian position coordinates, [m] \\
$X_j$ &$j$th component of vector $\X$ \\
$\X$ &generalized $n$-dimensional local or global Cartesian coordinates \\
$z_c$ &charge per mass of species $c$, [C kg\sub{c}\sup{-1}] \\
$\loc{z}$ &local specific charge, [C kg\sup{-1}] \\
$\ins{z}$ &velocity-distinct specific charge, [C kg\sup{-1}] \\
$\ins{\loc{z}}$ &velocity-distinct local specific charge, [C kg\sup{-1}] \\
$Z$ &total charge, [C] \\
\hline
\multicolumn{2}{l}{\bf Greek symbols}\\
$\alpha = \rho \, \loc{\alpha}$ &generalized volumetric density, [qty m\sup{-3}] \\
$\loc{\alpha}$ &local generalized specific density, [qty kg\sup{-1}] \\
$\beta = \de  \, \ins{\beta}$ &generalized velocimetric density, [qty (m s\sup{-1})\sup{-3}] \\
$\ins{\beta}$ &velocity-distinct generalized specific density, [qty kg\sup{-1}] \\
$\vGamma := \nabla_{\u} \x$ &inverse velocity gradient tensor field, [m (m s\sup{-1})\sup{-1}] = [s] \\
$\wGamma := \delf_{\u} \x$ &augmented inverse velocity gradient tensor field, components [m (m s\sup{-1})\sup{-1}] or [m s\sup{-1}] 
\\
$\vec{\delta}$ &Kronecker delta tensor \\
$\epsilon = \xi  \, \loc{\ins{\epsilon}}$ &generalized conditional volumetric density, [qty m\sup{-3}] \\
$\loc{\ins{\epsilon}}$ &velocity-distinct local generalized specific density, [qty kg\sup{-1}] \\
$\zeta$ &phase space fluid mass density,  [kg m\sup{-3} (m s\sup{-1})\sup{-3}] \\
$\zeta_c$ &phase space mass density of species $c$,  [kg\sub{c} m\sup{-3} (m s\sup{-1})\sup{-3}]  \\
$\eta$ &conditional velocimetric fluid mass density, [kg (m s\sup{-1})\sup{-3}] \\
$\eta_c$ &conditional velocimetric mass density of species $c$,  [kg\sub{c} (m s\sup{-1})\sup{-3}]  \\
$\theta = \eta \, \loc{\ins{\theta}}$ &generalized conditional velocimetric density, [qty (m s\sup{-1})\sup{-3}] \\
$\loc{\ins{\theta}}$ &velocity-distinct local generalized specific density, [qty kg\sup{-1}] \\
$\xi$ &conditional volumetric fluid mass density, [kg m\sup{-3}] \\
$\xi_c$ &conditional volumetric mass density of species $c$,  [kg\sub{c} m\sup{-3}]  \\
$\hat{\dot{\xi}}_c$ &molar rate of production of species $c$, [mol m\sup{-3} s\sup{-1}] \\
$\rho$ &volumetric fluid mass density, [kg m\sup{-3}] \\
$\rho_c$ &volumetric mass density of species $c$, [kg\sub{c} m\sup{-3}] \\
$\dot{\sigma}$ &total entropy production, [J K\sup{-1} s\sup{-1}]  \\
$\hat{\dot{\sigma}}$ &local entropy production, [J K\sup{-1} m\sup{-3} s\sup{-1}] \\
$\vec{\tau}$ &stress tensor (positive in compression), [Pa] = [J m\sup{-3}] \\
$\phi^{\C}$ &multivariate flow generated by $\V$ \\
$\hat{\phi}^{\C}$ &augmented multivariate flow generated by $\V \comp \C$ \\
$\varphi = \zeta  \, \loc{\ins{\varphi}} $ &generalized phase space density, [qty m\sup{-3} (m s\sup{-1})\sup{-3}] \\
$\loc{\ins{\varphi}}$ &velocity-distinct local generalized specific density, [qty kg\sup{-1}] \\
$\loc{\chi}_c$ &local specific mass density of species $c$, [kg\sub{c} kg\sup{-1}] \\
$\ins{\chi}_c$ &velocity-distinct specific mass density of species $c$, [kg\sub{c} kg\sup{-1}] \\
$\ins{\loc{\chi}}_c$ &velocity-distinct local specific mass density of species $c$, [kg\sub{c} kg\sup{-1}] \\
$\psi$ &generalized density of conserved quantity in generalized space \\
$\omega^r, \omega^n$ &$r$-form, $n$-form (respectively) in submanifold $\Omega$ \\
$\Omega$ &volumetric space occupied by a fluid (fluid volume), or general domain, or submanifold \\
\hline
\multicolumn{2}{l}{\bf Cyrillic symbols}\\
$\de$ &velocimetric fluid mass density, [kg (m s\sup{-1})\sup{-3}] \\
$\de_c$ &velocimetric mass density of species $c$, [kg\sub{c} (m s\sup{-1})\sup{-3}] \\
\hline\hline
\end{longtable}

 
 

\renewcommand\thetable{\arabic{table}}    
\setcounter{table}{0}    
\begin{turnpage}
\renewcommand{\arraystretch}{1.1}

\begin{table*}[!pt] \scriptsize \sffamily
\begin{center}
\caption{\label{tab:Re_tr_vol-temp-alpha_table}Conservation Laws for the Volumetric-Temporal Formulation (based on the Volumetric Fluid Density $\rho(\x,t)$).}
\begin{tabular*}{630pt}[t]{| L{47pt} | L{33pt} | R{95pt} | L{115pt} | C{45pt} | R{120pt} | L{95pt} | C{45pt} |}  
\hline\hline
{Conserved quantity}
& 
{Density}
&
\multicolumn{3}{l |}{{Integral Equation}}
&
\multicolumn{3}{l |}{{Differential Equation}}
\\
\hline
&
$\alpha(\x,t)$ $=\rho \loc{\alpha}$
&
LHS = $\dfrac{dQ}{dt}$ = $\dfrac{DQ}{Dt}$
&
RHS
&
SI Units
&
LHS
&
RHS
&
SI Units
\\
\hline \hline
{Fluid mass}
&
$\rho$
&
$0=\dot{M}_{FV}$ $=\dfrac{d \lvol \rho \rvol}{dt}$ $=\frac{d}{dt} \iiint\limits_{\Omega(t)} \rho \, \dV $
&
$= \iiint\limits_{\Omega(t)}  \biggl[ \dfrac{\partial \rho }{\partial t}   +  \nabla_{\x} \cdot (\rho  \u ) \biggr ] \dV $
&
[kg s\sup{-1}]
&
$0$
&
$= \dfrac{\partial \rho }{\partial t}   +  \nabla_{\x} \cdot (\rho  \u )  $
&
[kg s\sup{-1} m\sup{-3}]
\\
\hline
{Species mass}
&
$\rho \loc{\chi}_c$ $=\rho_c$ 
&
$(\dot{M}_c)_{FV} = \dfrac{d \lvol \rho \loc{\chi}_c \rvol}{dt}$
&
$ = \iiint\limits_{\Omega(t)}  \biggl[ \dfrac{\partial \rho \loc{\chi}_c }{\partial t} +  \nabla_{\x} \cdot (\rho \loc{\chi}_c  \u ) \biggr ] \dV$ 
&
[kg\sub{c} s\sup{-1}]
&
$
 \rho \dfrac{D \loc{\chi}_c}{Dt} 
=  M_c( \hat{\dot{\xi}}_c - \nabla_{\x} \cdot \vec{j}_c )$
&
$= \dfrac{\partial \rho \loc{\chi}_c}{\partial t}$ \mbox{$+ \nabla_{\x} \cdot (\rho \loc{\chi}_c  \u )$}
&
[kg\sub{c} s\sup{-1} m\sup{-3}]
\\
\hline
{Linear} \mbox{{momentum}}
&
$\rho \u$ 
&
$\sum \vec{F}_{FV} = \dfrac{d \lvol \rho \u \rvol}{dt} $
&
$ = \iiint\limits_{\Omega(t)}  \biggl[ \dfrac{\partial \rho \u }{\partial t}   +  {\nabla_{\x} \cdot (\rho\u  \u )} \biggr ] \dV $
&
[(kg m s\sup{-1}) s\sup{-1} = N]
&
$\rho \dfrac{D \u}{Dt} 
= -\nabla_{\x} P - \nabla_{\x} \cdot \vec{\tau} +  \rho \vec{g}$
&
$ = \dfrac{\partial \rho \u }{\partial t}   +  \nabla_{\x} \cdot (\rho\u  \u )$
&
[N m\sup{-3}]
\\
\hline\hline
\end{tabular*} 
\begin{tabular*}{653pt}[t]{ L{50pt} L{603pt} }  
\multicolumn{2}{l}{$^a$ It can be shown\cite{White_1986, Hutter_Johnk_2004} that the differential equation for angular momentum reduces to $\vec{\tau}=\vec{\tau}^\top$.}
\end{tabular*} 
\end{center}
\end{table*}

\begin{table*}[p] \scriptsize \sffamily
\begin{center}
\caption{\label{tab:Re_tr_vel-temp-beta_table}Conservation Laws for the Velocimetric-Temporal Formulation (based on the Velocimetric Fluid Density $\de(\u,t)$).}
\begin{tabular*}{630pt}[t]{| L{45pt} | L{35pt} | R{95pt} | L{160pt} | C{45pt} | R{40pt} | L{130pt} | C{45pt} |}  
\hline\hline
{Conserved quantity}
& 
{Density}
&
\multicolumn{3}{l |}{{Integral Equation}}
&
\multicolumn{3}{l |}{{Differential Equation}}
\\
\hline
&
$\beta(\u,t)$ $=\de \ins{\beta}$
&
LHS = $\dfrac{dQ}{dt}$ = $\dfrac{DQ}{Dt}$
&
RHS
&
SI Units
&
LHS
&
RHS
&
SI Units
\\
\hline \hline
{Fluid mass}
&
$\de$
&
$0=\dot{M}_{VV}$ $=\dfrac{d \lvel \de \rvel}{dt}$ $=\frac{d}{dt} \iiint\limits_{\D(t)} \de \, \dU $
&
$= \iiint\limits_{\D(t)}  \biggl[ \dfrac{\partial \de }{\partial t}   +  \nabla_{\u} \cdot (\de  \dot{\u} ) \biggr ] \dU $
&
[kg s\sup{-1}]
&
$0$
&
$= \dfrac{\partial \de }{\partial t}   +  \nabla_{\u} \cdot (\de   \dot{\u} )  $
&
[kg s\sup{-1} (m s\sup{-1})\sup{-3}]
\\
\hline
{Species mass}
&
$\de \ins{\chi}_c $ = $\de_c$ 
&
$(\dot{M}_c)_{VV} = \dfrac{d \lvel \de \ins{\chi}_c \rvel}{dt}$
&
$ = \iiint\limits_{\D(t)}  \biggl[ \dfrac{\partial \de \ins{\chi}_c }{\partial t}   +  \nabla_{\u} \cdot (\de \ins{\chi}_c   \dot{\u} ) \biggr ] \dU $ 
&
[kg\sub{c} s\sup{-1}]
&
$\de \dfrac{d \ins{\chi}_c}{dt}$
&
$=  \dfrac{\partial \de \ins{\chi}_c }{\partial t}   +  \nabla_{\u} \cdot (\de \ins{\chi}_c   \dot{\u} ) 
$
&
[kg\sub{c} s\sup{-1} (m s\sup{-1})\sup{-3}]
\\
\hline
{Linear \mbox{momentum}} 
&
$\de \u$
&
$\sum \vec{F}_{VV} = \dfrac{d \lvel \de \u \rvel}{dt} $
&
$ = \iiint\limits_{\D(t)}  \biggl[ \dfrac{\partial \de \u }{\partial t}  +  {\nabla_{\u} \cdot (\de \u   \dot{\u} )} \biggr ] \dU $
&
[(kg m s\sup{-1}) s\sup{-1} = N]
&
$\de \dfrac{d \u}{dt}$
&
$ = \dfrac{\partial \de \u }{\partial t}  +  {\nabla_{\u} \cdot (\de\u   \dot{\u} )}$
&
[N (m s\sup{-1})\sup{-3}]
\\
\hline\hline
\end{tabular*} 
\end{center}
\end{table*}


\begin{table*}[p] \scriptsize \sffamily
\begin{center}
\caption{\label{tab:Re_tr_velvol-temp_table}Conservation Laws for the Velocivolumetric-Temporal Formulation (based on the Phase Space Fluid Density $\zeta(\u,\x,t)$).}
\begin{tabular*}{630pt}[t]{| L{45pt} | L{35pt} | R{95pt} | L{160pt} | C{45pt} | R{40pt} | L{130pt} | C{45pt} |}  
\hline\hline
{Conserved quantity}
& 
{Density}
&
\multicolumn{3}{l |}{{Integral Equation}}
&
\multicolumn{3}{l |}{{Differential Equation}}
\\
\hline
&
$\varphi(\u,\x,t)$ $=\zeta \loc{\ins{\varphi}}$
&
LHS = $\dfrac{dQ}{dt}$ = $\dfrac{DQ}{Dt}$
&
RHS
&
SI Units
&
LHS
&
RHS
&
SI Units
\\
\hline \hline
{Fluid mass}
&
$\zeta $
&
$0=\dot{M}_{PV}$ $=\dfrac{d \, \lvol \lvel \zeta \rvel \rvol}{dt}$ $=\frac{d}{dt} \iiint\limits_{\Omega(t)} \iiint\limits_{\D(\x,t)} \zeta \, \dU \dV $
&
$= \iiint\limits_{\Omega(t)} \iiint\limits_{\D(\x,t)}  \biggl[ \dfrac{\partial \zeta }{\partial t}   
+  {\nabla_{\u} \cdot ( \zeta    \dot{\u} )} 
$ $+  {\nabla_{\x} \cdot ( \zeta   \u  )}
 \biggr ] \dU \dV $
&
[kg s\sup{-1}]
&
$0$
&
$= \dfrac{\partial \zeta }{\partial t}   
{+  \nabla_{\u} \cdot( \zeta    \dot{\u} )}  
$ ${+  \nabla_{\x} \cdot ( \zeta    \u  )}  $
&
[kg s\sup{-1} m\sup{-3} \mbox{(m s\sup{-1})\sup{-3}}]
\\
\hline
{Species mass}
&
$\zeta \loc{\ins{\chi}}_c $ = $\zeta_c$ 
&
$(\dot{M}_c)_{PV} = \dfrac{d \, \lvol \lvel \zeta \loc{\ins{\chi}}_c \rvel \rvol}{dt}$
&
$ = \iiint\limits_{\Omega(t)} \iiint\limits_{\D(\x,t)}  \biggl[ \dfrac{\partial \zeta \loc{\ins{\chi}}_c }{\partial t}   
+  {\nabla_{\u} \cdot ( \zeta \loc{\ins{\chi}}_c \dot{\u}  )}
$ $+ { \nabla_{\x} \cdot ( \zeta \loc{\ins{\chi}}_c \u  )}
 \biggr ] \dU \dV $ 
&
[kg\sub{c} s\sup{-1}]
&
$\zeta \dfrac{d \loc{\ins{\chi}}_c}{dt}$
&
$=  \dfrac{\partial \zeta \loc{\ins{\chi}}_c }{\partial t}   
+  {\nabla_{\u} \cdot ( \zeta \loc{\ins{\chi}}_c \dot{\u} )} 
$ $+  {\nabla_{\x} \cdot ( \zeta \loc{\ins{\chi}}_c \u )} $
&
[kg\sub{c} s\sup{-1} m\sup{-3} \mbox{(m s\sup{-1})\sup{-3}}]
\\
\hline\hline
\end{tabular*} 
\end{center}
\end{table*}


\begin{table*}[p] \scriptsize \sffamily
\begin{center}
\caption{\label{tab:Re_tr_vel-temp-varphi_table}Conservation Laws for the Velocimetric-Temporal Formulation (based on the Phase Space Fluid Density $\zeta(\u,\x,t)$).}
\begin{tabular*}{630pt}[t]{| L{45pt} | L{33pt} | R{73pt} | L{160pt} | C{44pt} | R{65pt} | L{130pt} | C{44pt} |}  
\hline\hline
{Conserved quantity}
& 
{Density}
&
\multicolumn{3}{l |}{{Integral Equation}}
&
\multicolumn{3}{l |}{{Differential Equation}}
\\
\hline
&
$\varphi(\u,\x,t)$ $=\zeta \loc{\ins{\varphi}}$
&
LHS = ${\dfrac{\partial \alpha}{\partial t} = \dfrac{\partial  (\rho \loc{\alpha})}{\partial t}}$
&
RHS
&
SI Units
&
LHS
&
RHS
&
SI Units
\\
\hline \hline
{Fluid mass}
&
$\zeta $
&
$\dfrac{\partial \rho}{\partial t}$ $=\dfrac{\partial  \, \lvel \zeta \rvel }{\partial t}$ $=\dfrac{\partial }{\partial t}  \iiint\limits_{\D(\x,t)} \zeta \, \dU  $
&
$=  \iiint\limits_{\D(\x,t)}  \biggl[ \dfrac{\partial \zeta }{\partial t}  { +  \nabla_{\u} \cdot ( \zeta   \dot{\u}  ) }\biggr ] \dU $
&
[kg m\sup{-3} s\sup{-1}]
&
$\mathcal{L}_{\dot{\u} \comp t}^{(t)}(\zeta \vol^3_{\u})$
&
$= \biggl[ \dfrac{\partial \zeta }{\partial t}   {+  \nabla_{\u} \cdot ( \zeta   \dot{\u} ) } \biggr] \dU  $
&
[kg m\sup{-3} s\sup{-1}]
\\
\hline
{Species mass}
&
$\zeta \loc{\ins{\chi}}_c $ = $\zeta_c$ 
&
$\dfrac{\partial  (\rho \loc{\chi}_c)}{\partial t}= \dfrac{\partial  \, \lvel \zeta \loc{\ins{\chi}}_c \rvel }{\partial t}$
&
$ =  \iiint\limits_{\D(\x,t)}  \biggl[ \dfrac{\partial \zeta \loc{\ins{\chi}}_c }{\partial t}   {+  \nabla_{\u} \cdot ( \zeta \loc{\ins{\chi}}_c   \dot{\u} )} \biggr ] \dU  $ 
&
[kg\sub{c} m\sup{-3} s\sup{-1}]
&
$\mathcal{L}_{\dot{\u} \comp t}^{(t)}(\zeta \loc{\ins{\chi}}_c \vol^3_{\u})$
&
$= \biggl[ \dfrac{\partial \zeta \loc{\ins{\chi}}_c }{\partial t}   {+  \nabla_{\u} \cdot ( \zeta \loc{\ins{\chi}}_c   \dot{\u} ) } \biggr] \dU
$
&
[kg\sub{c} m\sup{-3} s\sup{-1}]
\\
\hline\hline
\end{tabular*} 
\end{center}
\end{table*}


\begin{table*}[p] \scriptsize \sffamily
\begin{center}
\caption{\label{tab:Re_tr_vol-temp-varphi_table}Conservation Laws for the Volumetric-Temporal Formulation (based on the Phase Space Fluid Density $\zeta(\u,\x,t)$).}
\begin{tabular*}{630pt}[t]{| L{45pt} | L{35pt} | R{80pt} | L{150pt} | C{50pt} | R{65pt} | L{120pt} | C{50pt} |}  
\hline\hline
{Conserved quantity}
& 
{Density}
&
\multicolumn{3}{l |}{{Integral Equation}}
&
\multicolumn{3}{l |}{{Differential Equation}}
\\
\hline
&
$\varphi(\u,\x,t)$ $=\zeta \, \loc{\ins{\varphi}}$
&
LHS = ${\dfrac{\partial \beta}{\partial t}=\dfrac{\partial  (\de \ins{\beta})}{\partial t}}$
&
RHS
&
SI Units
&
LHS
&
RHS
&
SI Units
\\
\hline \hline
{Fluid mass}
&
$\zeta $
&
$\dfrac{\partial \de}{\partial t}$ $=\dfrac{\partial  \, \lvol \zeta \rvol }{\partial t}$ $=\dfrac{\partial }{\partial t}  \iiint\limits_{\Omega(\u,t)} \zeta \, \dV  $
&
$=  \iiint\limits_{\Omega(\u,t)}  \biggl[ \dfrac{\partial \zeta }{\partial t}   {+  \nabla_{\x} \cdot ( \zeta   \u  )} \biggr ] \dV  $
&
[kg {(m s\sup{-1})\sup{-3}} s\sup{-1}]
&
$\mathcal{L}_{\u \comp t}^{(t)}(\zeta  \vol^3_{\x})$
&
$= \biggl[ \dfrac{\partial \zeta }{\partial t}   {+  \nabla_{\x} \cdot ( \zeta   \u )} \biggr] \dV  $
&
[kg {(m s\sup{-1})\sup{-3}} s\sup{-1}]
\\
\hline
{Species mass}
&
$\zeta \loc{\ins{\chi}}_c $ = $\zeta_c$ 
&
$\dfrac{\partial  (\de \ins{\chi}_c)}{\partial t}= \dfrac{\partial  \, \lvol \zeta \loc{\ins{\chi}}_c \rvol }{\partial t}$
&
$ =  \iiint\limits_{\Omega(\u,t)}  \biggl[ \dfrac{\partial \zeta \loc{\ins{\chi}}_c }{\partial t}   {+  \nabla_{\x} \cdot ( \zeta \loc{\ins{\chi}}_c   \u )} \biggr ] \dV  $ 
&
[kg\sub{c} {(m s\sup{-1})\sup{-3}} s\sup{-1}]
&
$\mathcal{L}_{\u \comp t}^{(t)}(\zeta \loc{\ins{\chi}}_c \vol^3_{\x})$
&
$= \biggl[ \dfrac{\partial \zeta \loc{\ins{\chi}}_c }{\partial t}   {+  \nabla_{\x} \cdot ( \zeta \loc{\ins{\chi}}_c   \u )} \biggr] \dV
$
&
[kg\sub{c} {(m s\sup{-1})\sup{-3}} s\sup{-1}]
\\
\hline\hline
\end{tabular*} 
\end{center}
\end{table*}



\begin{table*}[p] \scriptsize \sffamily
\begin{center}
\caption{\label{tab:Re_tr_vel-spat_table}Conservation Laws for the Velocimetric-Spatial (Time-Independent) Formulation (based on the Phase Space Fluid Density $\zeta(\u,\x)$).}
\begin{tabular*}{630pt}[t]{| L{45pt} | L{35pt} | R{75pt} | L{145pt} | C{50pt} | R{65pt} | L{130pt} | C{50pt} |}  
\hline\hline
{Conserved quantity}
& 
{Density}
&
\multicolumn{3}{l |}{{Integral Equation}}
&
\multicolumn{3}{l |}{{Differential Equation}}
\\
\hline
&
$\varphi(\u,\x) = \zeta \ins{\loc{\varphi}}$
&
LHS = $\nabla_{\x} \alpha$ $=\nabla_{\x} (\rho \loc{\alpha})$
&
RHS
&
SI Units
&
LHS
&
RHS
&
SI Units
\\
\hline \hline
{Fluid mass}
&
$\zeta$
&
$\nabla_{\x} \rho$ $=\nabla_{\x} \lvel \zeta \rvel$ $= \nabla_{\x} \iiint\limits_{\D(\x)} \zeta \, \dU $
&
$= \iiint\limits_{\D(\x)} \bigl( \nabla_{\x} \zeta + {\nabla_{\u} \cdot ( \zeta \, \G^\top )} \bigr) \dU $
&
[kg m\sup{-4} = kg m\sup{-3} m\sup{-1}]
&
$ \mathcal{L}_{\G^\top \comp \x}^{(\x)} (\zeta \vol^3_{\u})$
&
$= \bigl( \nabla_{\x} \zeta + {\nabla_{\u} \cdot ( \zeta \, \G^\top )} \bigr) \dU$
&
[kg m\sup{-4} = kg m\sup{-3} m\sup{-1}]
\\
\hline
{Species mass}
&
$\zeta \ins{\loc{\chi}}_c$ $=\zeta_c$ 
&
$\nabla_{\x} (\rho \loc{\chi}_c) = \nabla_{\x} (\rho_c)$ $= \nabla_{\x}  \lvel \zeta \ins{\loc{\chi}}_c \rvel $
&
$ = \iiint\limits_{\D(\x)} \bigl( \nabla_{\x} (\zeta \ins{\loc{\chi}}_c) + \nabla_{\u} \cdot ( \zeta \ins{\loc{\chi}}_c \, \G^\top ) \bigr) \dU $ 
&
[kg\sub{c} m\sup{-4} = kg\sub{c} m\sup{-3} m\sup{-1}]
&
$ \mathcal{L}_{\G^\top \comp \x}^{(\x)} (\zeta \ins{\loc{\chi}}_c \vol^3_{\u})$
&
$= \bigl( \nabla_{\x} (\zeta \ins{\loc{\chi}}_c) + {\nabla_{\u} \cdot ( \zeta \ins{\loc{\chi}}_c \, \G^\top )} \bigr) \dU  $ 
&
[kg\sub{c} m\sup{-4} = kg\sub{c} m\sup{-3} m\sup{-1}]
\\
\hline\hline
\end{tabular*} 
\end{center}
\end{table*}


\begin{table*}[p] \scriptsize \sffamily
\begin{center}
\caption{\label{tab:Re_tr_vol-vel_table}Conservation Laws for the Volumetric-Velocital (Time-Independent) Formulation (based on the Phase Space Fluid Density $\zeta(\u,\x)$).}
\begin{tabular*}{630pt}[t]{| L{45pt} | L{35pt} | R{80pt} | L{140pt} | C{55pt} | R{65pt} | L{120pt} | C{55pt} |}  
\hline\hline
{Conserved quantity}
& 
{Density}
&
\multicolumn{3}{l |}{{Integral Equation}}
&
\multicolumn{3}{l |}{{Differential Equation}}
\\
\hline
&
$\varphi(\u,\x)$ $= \zeta \, \ins{\loc{\varphi}}$
&
LHS = $\nabla_{\u} \beta$ $=\nabla_{\u} (\de \ins{\beta})$
&
RHS
&
SI Units
&
LHS
&
RHS
&
SI Units
\\
\hline \hline
{Fluid mass}
&
$\zeta$
&
$\nabla_{\u} \de$ $=\nabla_{\u} \lvol \zeta \rvol$ $= \nabla_{\u} \iiint\limits_{\Omega(\u)} \zeta \, \dV $
&
$= \iiint\limits_{\Omega(\u)} \bigl( \nabla_{\u} \zeta + \nabla_{\x} \cdot ( \zeta \, \vGamma^\top ) \bigr) \dV $
&
[kg (m s\sup{-1})\sup{-4} = kg \mbox{(m s\sup{-1})\sup{-3}} \mbox{(m s\sup{-1})\sup{-1}}]
&
$ \mathcal{L}_{\vGamma^\top \comp \u}^{(\u)} (\zeta \vol^3_{\x})$
&
$= \bigl( \nabla_{\u} \zeta + \nabla_{\x} \cdot ( \zeta \, \vGamma^\top ) \bigr) \dV$
&
[kg (m s\sup{-1})\sup{-4} = kg \mbox{(m s\sup{-1})\sup{-3}} \mbox{(m s\sup{-1})\sup{-1}}]
\\
\hline
{Species mass}
&
$\zeta \ins{\loc{\chi}}_c$ $=\zeta_c$ 
&
$\nabla_{\u} (\de \ins{\chi}_c)$ $=\nabla_{\u} (\de_c)$ $= \nabla_{\u} \lvol \zeta \ins{\loc{\chi}}_c \rvol $
&
$ = \iiint\limits_{\Omega(\u)} \bigl( \nabla_{\u} (\zeta \ins{\loc{\chi}}_c) + {\nabla_{\x} \cdot ( \zeta \ins{\loc{\chi}}_c \, \vGamma^\top )} \bigr) \dV $ 
&
[kg\sub{c} (m s\sup{-1})\sup{-4} = kg\sub{c} \mbox{(m s\sup{-1})\sup{-3}} \mbox{(m s\sup{-1})\sup{-1}}]
&
$ \mathcal{L}_{\vGamma^\top \comp \u}^{(\u)} (\zeta \ins{\loc{\chi}}_c \vol^3_{\x})$
&
$= \bigl( \nabla_{\u} (\zeta \ins{\loc{\chi}}_c) + {\nabla_{\x} \cdot ( \zeta \ins{\loc{\chi}}_c \, \vGamma^\top )} \bigr) \dV  $ 
&
[kg\sub{c} (m s\sup{-1})\sup{-4} = kg\sub{c} \mbox{(m s\sup{-1})\sup{-3}} \mbox{(m s\sup{-1})\sup{-1}}]
\\
\hline\hline
\end{tabular*} 
\end{center}
\end{table*}


\begin{table*}[p] \scriptsize \sffamily
\begin{center}
\caption{\label{tab:Re_tr_vel-spattemp_table}Conservation Laws for the Velocimetric-Spatiotemporal Formulation (based on the Phase Space Fluid Density $\zeta(\u,\x,t)$).}
\begin{tabular*}{630pt}[t]{| L{45pt} | L{35pt} | R{80pt} | L{145pt} | C{50pt} | R{70pt} | L{120pt} | C{50pt} |}  
\hline\hline
{Conserved quantity}
& 
{Density}
&
\multicolumn{3}{l |}{{Integral Equation}}
&
\multicolumn{3}{l |}{{Differential Equation}}
\\
\hline
&
$\varphi(\u,\x,t)$ $= \zeta \, \ins{\loc{\varphi}}$
&
LHS = $\delf_{\x} \alpha$ $=\delf_{\x} (\rho \loc{\alpha})$
&
RHS
&
SI Units
&
LHS
&
RHS
&
SI Units
\\
\hline \hline
{Fluid mass}
&
$\zeta$
&
$\delf_{\x} \rho$ $=\delf_{\x} \lvel \zeta \rvel$ $= \delf_{\x} \iiint\limits_{\D(\x,t)} \zeta \, \dU $
&
$= \iiint\limits_{\D(\x,t)} \bigl( \delf_{\x} \zeta + \nabla_{\u} \cdot ( \zeta \, \wG^\top) \bigr) \dU $
&
[kg m\sup{-3} $\vecunit_{\x,t}^{-1}$]
&
$ \mathcal{L}_{\wG^\top \comp \x,t}^{(\x, t)} (\zeta \vol^3_{\u})$
&
$= \bigl( \delf_{\x} \zeta + {\nabla_{\u} \cdot ( \zeta \, \wG^\top)} \bigr) \dU$
&
[kg m\sup{-3} $\vecunit_{\x,t}^{-1}$]
\\
\hline
{Species mass}
&
$\zeta \ins{\loc{\chi}}_c$ $=\zeta_c$ 
&
$\delf_{\x} (\rho \loc{\chi}_c)$ $= \delf_{\x} (\rho_c)$ $= \delf_{\x}  \lvel \zeta \ins{\loc{\chi}}_c \rvel $
&
$ = \iiint\limits_{\D(\x,t)} \bigl( \delf_{\x} (\zeta \ins{\loc{\chi}}_c) + {\nabla_{\u} \cdot ( \zeta \ins{\loc{\chi}}_c \, \wG^\top)} \bigr) \dU $ 
&
[kg\sub{c} m\sup{-3} $\vecunit_{\x,t}^{-1}$]
&
$ \mathcal{L}_{\wG^\top \comp \x,t}^{(\x, t)} (\zeta \ins{\loc{\chi}}_c \vol^3_{\u})$
&
$= \bigl( \delf_{\x} (\zeta \ins{\loc{\chi}}_c) + {\nabla_{\u} \cdot ( \zeta \ins{\loc{\chi}}_c \, \wG^\top)} \bigr) \dU  $ 
&
[kg\sub{c} m\sup{-3} $\vecunit_{\x,t}^{-1}$]
\\
\hline\hline
\end{tabular*} 
\end{center}
\end{table*}


\begin{table*}[p] \scriptsize \sffamily
\begin{center}
\caption{\label{tab:Re_tr_vol-veltemp_table}Conservation Laws for the Volumetric-Velocitotemporal Formulation (based on the Phase Space Fluid Density $\zeta(\u,\x,t)$).}

\begin{tabular*}{630pt}[t]{| L{45pt} | L{35pt} | R{75pt} | L{145pt} | C{50pt} | R{65pt} | L{130pt} | C{50pt} |}  
\hline\hline
{Conserved quantity}
& 
{Density}
&
\multicolumn{3}{l |}{{Integral Equation}}
&
\multicolumn{3}{l |}{{Differential Equation}}
\\
\hline
&
$\varphi(\u,\x,t) $ $= \zeta \, \ins{\loc{\varphi}}$
&
LHS = $\delf_{\u} \beta$ $=\delf_{\u} (\de \ins{\beta})$
&
RHS
&
SI Units
&
LHS
&
RHS
&
SI Units
\\
\hline \hline
{Fluid mass}
&
$\zeta$
&
$\delf_{\u} \de$ $=\delf_{\u} \lvol \zeta \rvol$ $= \delf_{\u} \iiint\limits_{\Omega(\u,t)} \zeta \, \dV $
&
$= \iiint\limits_{\Omega(\u,t)} \bigl( \delf_{\u} \zeta + \nabla_{\x} \cdot ( \zeta \, \wGamma^\top ) \bigr) \dV $
&
[kg (m s\sup{-1})\sup{-3} $\vecunit_{\u,t}^{-1}$]
&
$ \mathcal{L}_{\wGamma^\top \comp \u,t}^{(\u,t)} (\zeta \vol^3_{\x})$
&
$= \bigl( \delf_{\u} \zeta + \nabla_{\x} \cdot ( \zeta \, \wGamma^\top ) \bigr) \dV$
&
[kg (m s\sup{-1})\sup{-3} $\vecunit_{\u,t}^{-1}$]
\\
\hline
{Species mass}
&
$\zeta \ins{\loc{\chi}}_c$ $=\zeta_c$ 
&
$\delf_{\u} (\de \ins{\chi}_c) =\delf_{\u} (\de_c) $ $= \delf_{\u}  \lvol \zeta \ins{\loc{\chi}}_c \rvol $
&
$ = \iiint\limits_{\Omega(\u,t)} \bigl( \delf_{\u} (\zeta \ins{\loc{\chi}}_c) + \nabla_{\x} \cdot ( \zeta \ins{\loc{\chi}}_c \, \wGamma^\top ) \bigr) \dV $ 
&
[kg\sub{c} (m s\sup{-1})\sup{-3} $\vecunit_{\u,t}^{-1}$]
&
$ \mathcal{L}_{\wGamma^\top \comp \u,t}^{(\u,t)} (\zeta \ins{\loc{\chi}}_c \vol^3_{\x})$
&
$= \bigl( \delf_{\u} (\zeta \ins{\loc{\chi}}_c) + {\nabla_{\x} \cdot ( \zeta \ins{\loc{\chi}}_c \, \wGamma^\top )} \bigr) \dV  $ 
&
[kg\sub{c} (m s\sup{-1})\sup{-3} $\vecunit_{\u,t}^{-1}$]
\\
\hline\hline
\end{tabular*} 
\end{center}
\end{table*}

\begin{table*}[p] \scriptsize \sffamily
\begin{center}
\caption{\label{tab:Re_tr_vel-temp-theta_table}Conservation Laws for the Velocimetric-Temporal Formulation (based on the Velocimetric Fluid Density $\eta(\u,\x,t)$).}
\begin{tabular*}{630pt}[t]{| L{45pt} | L{35pt} | R{95pt} | L{160pt} | C{45pt} | R{40pt} | L{130pt} | C{45pt} |}  
\hline\hline
{Conserved quantity}
& 
{Density}
&
\multicolumn{3}{l |}{{Integral Equation}}
&
\multicolumn{3}{l |}{{Differential Equation}}
\\
\hline
&
$\theta(\u,\x,t)$ $=\eta \ins{\loc{\theta}}$
&
LHS = $\dfrac{dQ}{dt}$ = $\dfrac{DQ}{Dt}$
&
RHS
&
SI Units
&
LHS
&
RHS
&
SI Units
\\
\hline \hline
{Fluid mass}
&
$\eta$
&
$0=\dot{M}_{VV}$ $=\dfrac{d \lvel \eta \rvel}{dt}$ $=\frac{d}{dt} \iiint\limits_{\D(\x,t)} \eta \, \dU $
&
$= \iiint\limits_{\D(\x,t)}  \biggl[ \dfrac{\partial \eta }{\partial t}   +  {\nabla_{\u} \cdot (\eta  \dot{\u} )} \biggr ] \dU $
&
[kg s\sup{-1}]
&
$0$
&
$= \dfrac{\partial \eta }{\partial t}   +  {\nabla_{\u} \cdot (\eta   \dot{\u} )}  $
&
[kg s\sup{-1} (m s\sup{-1})\sup{-3}]
\\
\hline
{Species mass}
&
$\eta \ins{\loc{\chi}}_c $ = $\eta_c$ 
&
$(\dot{M}_c)_{VV} = \dfrac{d \lvel \eta \ins{\loc{\chi}}_c \rvel}{dt}$
&
$ = \iiint\limits_{\D(\x,t)}  \biggl[ \dfrac{\partial \eta \ins{\loc{\chi}}_c }{\partial t}   +  {\nabla_{\u} \cdot (\eta \ins{\loc{\chi}}_c   \dot{\u} )} \biggr ] \dU $ 
&
[kg\sub{c} s\sup{-1}]
&
$\eta \dfrac{d \ins{\loc{\chi}}_c}{dt}$
&
$=  \dfrac{\partial \eta \ins{\loc{\chi}}_c }{\partial t}   + { \nabla_{\u} \cdot (\eta \ins{\loc{\chi}}_c   \dot{\u} )} 
$
&
[kg\sub{c} s\sup{-1} (m s\sup{-1})\sup{-3}]
\\
\hline
{Linear \mbox{momentum}} 
&
$\eta \u$
&
$\sum \vec{F}_{VV} = \dfrac{d \lvel \eta \u \rvel}{dt} $
&
$ = \iiint\limits_{\D(\x,t)}  \biggl[ \dfrac{\partial \eta \u }{\partial t}  +  {\nabla_{\u} \cdot (\eta \u   \dot{\u} )} \biggr ] \dU $
&
[(kg m s\sup{-1}) s\sup{-1} = N]
&
$\eta \dfrac{d \u}{dt}$
&
$ = \dfrac{\partial \eta \u }{\partial t} +  {\nabla_{\u} \cdot (\eta\u   \dot{\u} )}$
&
[N (m s\sup{-1})\sup{-3}]
\\
\hline\hline
\end{tabular*} 
\end{center}
\end{table*}


\begin{table*}[p] \scriptsize \sffamily
\begin{center}
\caption{\label{tab:Re_tr_vol-temp-epsilon_table}Conservation Laws for the Volumetric-Temporal Formulation (based on the Volumetric Fluid Density $\xi(\u,\x,t)$).}
\begin{tabular*}{630pt}[t]{| L{45pt} | L{35pt} | R{95pt} | L{160pt} | C{45pt} | R{40pt} | L{130pt} | C{45pt} |}  
\hline\hline
{Conserved quantity}
& 
{Density}
&
\multicolumn{3}{l |}{{Integral Equation}}
&
\multicolumn{3}{l |}{{Differential Equation}}
\\
\hline
&
$\epsilon(\u,\x,t)$ $=\xi \ins{\loc{\epsilon}}$
&
LHS = $\dfrac{dQ}{dt}$ = $\dfrac{DQ}{Dt}$
&
RHS
&
SI Units
&
LHS
&
RHS
&
SI Units
\\
\hline \hline
{Fluid mass}
&
$\xi$
&
$0=\dot{M}_{FV}$ $=\dfrac{d \lvol \xi \rvol}{dt}$ $=\frac{d}{dt} \iiint\limits_{\Omega(\u,t)} \xi \, \dV $
&
$= \iiint\limits_{\Omega(\u,t)}  \biggl[ \dfrac{\partial \xi }{\partial t}   +  \nabla_{\x} \cdot (\xi  \u ) \biggr ] \dV $
&
[kg s\sup{-1}]
&
$0$
&
$= \dfrac{\partial \xi }{\partial t}   +  \nabla_{\x} \cdot (\xi  \u )  $
&
[kg s\sup{-1} m\sup{-3}]
\\
\hline
{Species mass}
&
$\xi \ins{\loc{\chi}}_c$ $=\xi_c$ 
&
$(\dot{M}_c)_{FV} = \dfrac{d \lvol \xi \ins{\loc{\chi}}_c \rvol}{dt}$
&
$ = \iiint\limits_{\Omega(\u,t)}  \biggl[ \dfrac{\partial \xi \ins{\loc{\chi}}_c }{\partial t} +  \nabla_{\x} \cdot (\xi \ins{\loc{\chi}}_c  \u ) \biggr ] \dV$ 
&
[kg\sub{c} s\sup{-1}]
&
$ \xi \dfrac{D \ins{\loc{\chi}}_c}{Dt} $
&
$= \dfrac{\partial \xi \ins{\loc{\chi}}_c}{\partial t}$ \mbox{$+ \nabla_{\x} \cdot (\xi \ins{\loc{\chi}}_c  \u )$}
&
[kg\sub{c} s\sup{-1} m\sup{-3}]
\\
\hline\hline
\end{tabular*} 
\end{center}
\end{table*}

\end{turnpage}


\begin{thebibliography}{99}


\bibitem
{Reynolds_1903} {O. Reynolds}, {``Papers on Mechanical and Physical Subjects''}, Vol. III, Cambridge Univ. Press, Cambridge UK (1903).


\bibitem
{White_1986} {F.M. White}, {``Fluid Mechanics''}, 2nd ed., McGraw-Hill Higher Education, NY (1986).

\bibitem
{Munson_etal_2010} {B.R. Munson, D.F. Young, T.H. Okiishi \& W.W. Huebsch}, {``Fundamentals of Fluid Mechanics''}, 6th ed., John Wiley, NY (2010).



\bibitem{deGroot_M_1984} S.R. de Groot \& P. Mazur, ``Non-Equilibrium Thermodynamics'', Dover Publ., NY (1984).

\bibitem{Bird_etal_2006} R.B. Bird, W.E. Stewart \& E.N. Lightfoot, ``Transport Phenomena'', 2nd ed., John Wiley \& Sons, NY (2006).

\bibitem
{White_2006}  {F.M. White}, {``Viscous Fluid Flow''}, 3rd ed., McGraw-Hill, NY (2006).

\bibitem{Durst_2008}
{F. Durst} {``Fluid Mechanics''} Springer-Verlag, Berlin (2008).


\bibitem
{Dvorkin_Goldschmidt_2005} {E.N. Dvorkin \& M.B. Goldschmit}, {``Nonlinear Continua''}, Springer-Verlag, Berlin (2006), chap. 4. 

\bibitem
{Truesdell_Toupin_1960} {C. Truesdell \& R.A. Toupin}, {``The classical field theories''}, {\it in} Fl\"ugge, S., Handbuch der Physik, Band III/1, Springer-Verlag, Berlin (1960), p. 347. 

\bibitem{Myers_2015} {M.K. Myers}, {``Generalized integral theorems and application to the equations of continuum mechanics''}, Aeroacoustics {\bf 14}(1-2), 1-24 (2015).

\bibitem
{Seguin_Fried_2014} B. Seguin \& E. Fried, {``Roughening it - Evolving irregular domains and transport theorem''}, {Mathematical Models and Methods in Applied Sciences} {\bf 24}(9),1729-1779  (2014). 

\bibitem
{Seguin_etal_2014} B. Seguin, D.F. Hinz \& E. Fried, {``Extending the transporttheorem to rough domains of integration''}, {Applied Mechanics Reviews} {\bf 66}, 050802 (2014). 

\bibitem
{Falach_Segev_2014} L. Falach \& R. Segev, {``Reynolds transport theorem for smooth deformations of currents on manifolds''}, {Mathematics and Mechanics of Solids} {\bf 20}(6), 770-786 (2015). 

\bibitem{Gurtin_etal_1989} M.E. {Gurtin}, A. {Struthers} \& W.O. {Williams}, {``A transport theorem for moving interfaces''}, Q Appl Math {\bf XLVIII}, 773-777 (1989).

\bibitem{Ochoa-Tapia_etal_1993} J.A. {Ochoa-Tapia}, J.A. {del Rio} \& S. {Whitaker}, {``Bulk and surface diffusion in porous media: an application of the surface-averaging theorem''}, Chem. Eng. Sci. {\bf 48}(11), 2061-2082  (1993). 

\bibitem{Slattery_Sagis_Oh_2007} J.C. {Slattery}, L. {Sagis} \& E.-S. {Oh}, {``Interfacial Transport Phenomena''}, 2nd ed., Springer, NY (2007).

\bibitem{Fosdick_Tang_2009} R. {Fosdick} \& H. {Tang}, {``Surface transport in continuum mechanics''}, Math Mech Solid {\bf 14}(6),  587-598 (2009).

\bibitem{Lidstrom_2011} P. {Lidstr\"om}, {``Moving regions in Euclidean space and Reynolds transport theorem''}, Mathematics and Mechanics of Solids {\bf 16}(4),  366-380 (2011).


\bibitem
{Flanders_1973} {H. Flanders}, {``Differentiation under the integral sign''}, {The American Mathematical Monthly} {\bf 80}(6), 615-627 (1973).

\bibitem{Lee_2009}
{J.M. Lee}, ``Manifolds and Differential Geometry'', American Mathematical Society, Providence RI USA (2009).

\bibitem{Frankel_2013}
{T. Frankel}, {``The Geometry of Physics''}, 3rd ed., Cambridge Univ. Press, UK (2013).

\bibitem
{Harrison_2015} {J. Harrison}, {``Operator calculus of differential chains and differential forms''}, {J. Geom. Anal.} {\bf 25},  357-420 (2015).


\bibitem
{Tai_1992} {C.-T. Tai} 1992 , {``Generalized Vector and Dyadic Analysis''}, IEEE, NY (1992).




\bibitem{Niven_etal_2020} R.K. Niven, L. Cordier, E. Kaiser, M. Schlegel, B.R. Noack, ``Rethinking the Reynolds transport theorem, Liouville equation, and Perron-Frobenius and Koopman operators'', arXiv:1810.06022v5 (2020).

\bibitem{Niven_etal_2020b} R.K. Niven, L. Cordier, E. Kaiser, M. Schlegel, B.R. Noack, ``New conservation laws based on generalised Reynolds transport theorems'', paper 110, in Chanson, H. \& Brown, R. (eds.), Proc. 22nd Australasian Fluid Mechanics Conference (AFMC2020), Brisbane, Australia, 7-10 Dec 2020, University of Queensland (2020), https://espace.library.uq.edu.au/view/UQ:380a993.


\bibitem
{Liouville_1838} J. Liouville, {``Note sur la th\'eorie de la variation des constantes arbitraires''}, {Journal de Math\'ematiques Pures et Appliqu\'ees} {\bf 1}(3), 342-349 (1838) (French).

\bibitem
{Risken_1984}  H. Risken, {``The Fokker-Planck Equation: Methods of Solution and Applications''}, Springer-Verlag, Berlin (1984).

\bibitem
{Lutzen_1990} J. L\"utzen, {``Joseph Liouville 1809-1882, Master of Pure and Applied Mathematics''}, Springer-Verlag, NY (1990), pp 637-708.

\bibitem
{Ehrendorfer_2003} M. Ehrendorfer, {``The Liouville equation in atmospheric predictability''}, Seminar on Predictability of Weather and Climate, 9-13 Sept. 2002, European Centre for Medium-Range Weather Forecasts (ECMWF), Reading, UK  (2003), pp 47-81.

\bibitem
{Pottier_2010} {N. Pottier}, {``Nonequilibrium Statistical Physics: Linear Irreversible Processes''},  Oxford Univ. Press, UK (2010).


\bibitem
{Feller_1966} {W. Feller}, {``An Introduction to Probability Theory and Its Applications''}, Volume II, John Wiley and Sons, Inc., NY, (1966).


\bibitem
{Batchelor_1967}  {G.K. Batchelor}, {``Theory of homogeneous turbulence''}, Cambridge Univ. Press, Cambridge UK (1967).

\bibitem
{Monin_Yaglom_1971a}  A.S. Monin \& A.M. Yaglom, {``Statistical Fluid Mechanics: Mechanics of Turbulence''}, Vol. I, Dover Publ., NY (1971).

\bibitem
{Hinze_1975}  {J.O. Hinze}, {``Turbulence: An Introduction to its Mechanism and Theory''}, 2nd ed.. McGraw-Hill, NY (1975).

\bibitem
{Pope_2000} {S.B. Pope}, {``Turbulent Flows''}, Cambridge Univ. Press, UK (2000).

\bibitem{Davidson_2004} {P.A. Davidson}, ``Turbulence: An Introduction for Scientists and Engineers'', Oxford Univ. Press, Oxford UK (2004).



\bibitem{Jaynes_1980} E.T. Jaynes {``The minimum entropy production principle''}, Ann. Rev. Phys. Chem. {\bf 31}, 579-601 (1980).

\bibitem{Gonzalez_etal_2013} D. Gonz\'alez, S. Davis \& G. Guti\'errez, ``Newtonian dynamics from the principle of maximum caliber'', arXiv:1310.1382v1 (2013).

\bibitem{Dixit_etal_2017} P.D. Dixit, J. Wagoner, C. Weistuch, S. Press\'e, K Ghosh, K.A. Dill, ``Maximum caliber: a general variational principle for dynamical systems'', arXiv:1711.03450v1 (2017).

\bibitem{Ghosh_etal_2020} K. Ghosh, P.D. Dixit, L. Agozzino, K.A. Dill, ``The maximum caliber variational principle for nonequilibria'', Annual Review of Physical Chemistry, {\bf 71}, 213-238 (2020).



\bibitem{Matheron_1965} G. Matheron, ``Les variables r\'egionalis\'ees et leur estimation, une application de la th\'eorie de fonctions al\'eatoires aux sciences de la nature'', Masson et Cie, Paris (1965). 

\bibitem{Marle_1967} {C.M. Marle}, ``\'Ecoulements monophasiques en milieu poreux'', Rev. Inst. Fran\c{c}ais du P\'etrole. {\bf 22}(10), 1471-1509 (1967). 

\bibitem{Anderson_Jackson_1967} {T.B. Anderson \& R. Jackson}, ``A fluid mechanical description of fluidized beds'', Ind. Engng Chem. Fundam. {\bf 6}, 527-539 (1967). 

\bibitem{Whitaker_1967a} {S. Whitaker}, {``Diffusion and dispersion in porous media''}, AIChE J. {\bf 13}(3), 420-427 (1967). 

\bibitem{Whitaker_1969} {S. Whitaker}, {``Advances in theory of fluid motion in porous media''}, Ind. Eng. Chem. {\bf 61}, 14-28 (1969). 

\bibitem{Bachmat_1972} {Y. Bachmat}, {``Spatial macroscopization of processes in heterogeneous systems''}, Israel Journal of Technology {\bf 10}(5), 391-403 (1972). 

\bibitem{Gray_Lee_1977} {W.G. Gray \& P.C.Y. Lee}, ``On the theorems for local volume averaging of multiphase systems'', Int. J. Multiphase Flow {\bf 3}, 333-340 (1977). 

\bibitem{Hassanizadeh_Gray_1979a} M. Hassanizadeh \& W.G. Gray, ``General conservation equations for multi-phase systems: 1. Averaging procedure'', Adv. Water Resources {\bf 2}, 131-144 and {\bf 3}, 91-94 (1979). 

\bibitem{Narasimhan_1980} T.N. Narasimhan, ``A note on volume-averaging'', Adv. Water Res. {\bf 3}, 135-139 (1980). 

\bibitem{Ene_1981} H.L. Ene, ``On the thermodynamic theory of mixtures'', Int. J. Eng. Sci. {\bf 19}, 905-914 (1981). 

\bibitem{Cushman_1982} {J.H. Cushman}, ``Proofs of the volume averaging theorems for multiphase flow'', Adv Water Resources {\bf 5}, 248-253 (1982). 

\bibitem{Cushman_1983b} {J.H. Cushman}, ``Volume averaging, probabilistic averaging, and ergodicity'', Adv. Water Resources {\bf 6}, 182-184 (1983). 

\bibitem{Baveye_Sposito_1984} P. Baveye \& G. Sposito, ``The operational significance of the continuum hypothesis in the theory of water movement through soils and aquifers'', Water Resources Research {\bf 20}(5), 521-530 (1984). 

\bibitem{Cushman_1984} J.H. Cushman, ``On unifying the concepts of scale, instrumentation, and stochastics in the development of multiphase transport theory'', Water Res. Res. {\bf 20}(11), 1668-1676 (1984). 

\bibitem{Cushman_1985} J.H. Cushman, ``Multiphase transport based on compact distributions'', Acta Applicandae Mathematicae {\bf 3}, 239-254 (1985). 

\bibitem{Bear_Bachmat_1991} {J. Bear \& Y. Bachmat}, {``Introduction to Modeling of Transport Phenomena in Porous Media''}, Kluwer Academic Publ., Dordrecht, Netherlands (1991). 

\bibitem{Gray_etal_1993} {W.G. Gray, A. Leijnse, R.L. Kolar, C.A. Blain}, ``Mathematical Tools for Changing Scale in the Analysis of Physical Systems'', CRC Press, Bosa Roca, USA (1993). 

\bibitem{Quintard_Whitaker_1993} M. Quintard \& S. Whitaker, ``Transport in ordered and disordered porous media: volume-averaged equations, closure problems, and comparison with experiment'', Chem. Eng. Sci. {\bf 48}(14), 2537-2564 (1993). 

\bibitem{Chen_1994} {Z. Chen}, ``Large-scale averaging analysis of single phase flow in fractured reservoirs'', SIAM J. Appl. Math. {\bf 54}(3), 641-659 (1994). 

\bibitem{Grau_Cantero_1994} {R.J. Grau \& H.J. Cantero}, {``A systematic time-, space- and time-space-averaging procedure for bulk phase equations in systems with multiphase flow''}, Chem. Eng. Sci. {\bf 49}(4), 449-461 (1994). 

\bibitem{Cushman_1997} J.H. Cushman, ``The Physics of Fluids in Hierarchical Porous Media: Angstroms to Miles'', Springer, NY (1997). 

\bibitem{Whitaker_1999}
{S. Whitaker}, {``The Method of Volume Averaging''}, Kluwer Academic Publishers, Netherlands (1999). 

\bibitem{Gray_Miller_2005b} W.G. Gray \& C.T. Miller, ``Thermodynamically constrained averaging theory approach for modeling flow and transport phenomena in porous medium systems: 2. Foundation'', Adv. Water Resources {\bf 28}, 181-202 (2005). 

\bibitem{Wood_2013} {W.D. Wood}, {``Technical note: Revisiting the geometric theorems for volume averaging''}, Advances in Water Resources {\bf 62}, 340-352 (2013). 

\bibitem{Gray_Miller_2013} W.G. Gray \& C.T. Miller, ``A generalization of averaging theorems for porous medium analysis'', Advances in Water Resources {\bf 62}, 227-237 (2013). 

\bibitem{Pokrajac_deLemos_2015}
{P. Pokrajac \& M.J.S. de Lemos}, {``Spatial averaging over a variable volume and Its application to boundary-layer flows over permeable walls''}, J. Hydraul. Eng. {\bf 141}(4), 04014087 (2015). 

\bibitem{Davit_Quintard_2017} Y. Davit \& M. Quintard, ``Technical notes on volume averaging in porous media I: How to choose a spatial averaging operator for periodic and quasiperiodic structures'', Transp Porous Med. {\bf 119}, 555-584 (2017). 

\bibitem{Takatsu_2017}
{Y. Takatsu}, {``Modification of the fundamental theorem for transport phenomena in porous media''}, Int. J. Heat and Mass Transfer {\bf 115}, 1109-1120 (2017). 



\bibitem{Kobayashi_Nomizu_1963}
{S. Kobayashi \& K. Nomizu}, {``Foundations of Differential Geometry''} Vol. 1, John Wiley \& Sons, NY, Interscience Publ. (1963).

\bibitem{Guggenheimer_1963}
{H.W. Guggenheimer} {``Differential Geometry''}, McGraw-Hill, NY (1963), also Dover Publ., NY, (1977).

\bibitem
{Cartan_1970} {H. Cartan} ``Differential Forms'', Dover Publ., NY (1970).


\bibitem{Lovelock_Rund_1989}
{D. Lovelock \& H. Rund}, {``Tensors, Differential Forms, and Variational Principles''},  Dover Publ., NY (1989).

\bibitem{Olver_1993}
{P.J. Olver}, {``Applications of Lie Groups to Differential Equations''}, 2nd ed., Springer NY (1993).


\bibitem{Torres_del_Castillo_2012}
{G. Torres del Castillo}, {``Differentiable Manifolds''}, Springer NY (2012).

\bibitem{Bachman_2012}
{D. Bachman}, {``A Geometric Approach to Differential Forms''}, 2nd ed., Birkh\"auser, Springer, NY (2012).


\bibitem{Sjamaar_2017}
{R. Sjamaar}, {``Manifolds and Differential Forms''}, Cornell University, NY (2017).


\bibitem{Spurk_1997}
J.H. Spurk, {``Fluid Mechanics''}, Berlin, Springer (1997).





\bibitem{Kuiken_1994} G.D.C. Kuiken, ``Thermodynamics of Irreversible Processes'', John Wiley \& Sons, Chichester (1994).

\bibitem{Demirel_2002} Y. Demirel, ``Nonequilibrium Thermodynamics'', Elsevier, NY (2002). 

\bibitem{Hutter_Johnk_2004} K. Hutter \& K.D. J\"ohnk, ``Continuum Methods of Physical Modeling'', Springer, NY (2004). 



\bibitem{Weinstock_1952}
{R. Weinstock}, {``Calculus of Variations''}, Dover Publ., Mineola, NY (1952).

\bibitem{Gelfand_Fomin_1963} 
{I.M. Gelfand \& S.V. Fomin}, {``Calculus of Variations''}, Dover Publ., Mineola, NY (1963).



\bibitem{Truesdell_1954} C. Truesdell, ``The Kinematics of Vorticity'', Dover Publ., Mineola, NY (2018).



\bibitem{Cox_1961} R.T. Cox, ``The Algebra of Probable Inference'', John Hopkins Press, NJ (1961).

\bibitem{Jaynes_2003} E.T. Jaynes (G.L. Bretthorst, ed.) {``Probability Theory: The Logic of Science''}, Cambridge Univ. Press, UK (2003).


\end{thebibliography}
\end{document}